\documentclass[12pt]{article}
\pdfoutput=1

\usepackage[left=0.8in,right=0.8in,top=1.0in,bottom=1.0in]{geometry}
\usepackage{amsmath,amssymb,cite,slashed,tabularx}
\usepackage{subfigure}
\usepackage[dvipdfmx]{graphicx,color}
\usepackage{ascmac}
\usepackage[T1]{fontenc}
\usepackage{textcomp}

\usepackage{ulem}
\usepackage[sectionbib]{chapterbib}
\usepackage{graphicx}
\usepackage{cancel}
\usepackage{ulem}
\usepackage{multirow}
\usepackage{empheq,fancybox}
\usepackage{lscape}
\usepackage{cancel}
\usepackage{bm}
\usepackage{amssymb}
\usepackage{arydshln}
\usepackage{adjustbox}
\usepackage{mathtools}

\allowdisplaybreaks
\usepackage{caption}
\definecolor{amber}{rgb}{1.0, 0.49, 0.0}
\definecolor{green}{rgb}{0.0, 0.5, 0.0}
\definecolor{applegreen}{rgb}{0.55, 0.71, 0.0}
\definecolor{pink}{rgb}{0.94, 0.5, 0.5}
\definecolor{lightblue}{rgb}{0.39, 0.58, 0.93} 
\definecolor{purple}{rgb}{0.59, 0.44, 0.84} 
\definecolor{bluencs}{rgb}{0.0, 0.53, 0.74} 

\usepackage[
bookmarkstype=toc, colorlinks=true, linkcolor=blue, citecolor=green, pdfborder={0 0 0}, bookmarks=true, bookmarksnumbered=true]{hyperref}

\newcommand{\CL}[2]{\mathcal{C}_{\underset{#2}{#1}}}

\newcommand{\brackets}[1]{\left( #1 \right)}

\def\Slash#1{\rlap/#1}
\def\Slash#1{\rlap{\kern .1em /}#1}

\usepackage{here}
\usepackage {cancel}

\begin{document}
	
	
	\title{
		\begin{flushright}
			\ \\*[-80pt]
			\begin{minipage}{0.2\linewidth}
				\normalsize
			\end{minipage}
		\end{flushright}
	%
		{\Large \bf
		Electron EDM and  LFV decays in the light of
		\\*[20pt]
		Muon $(g-2)_\mu$ with  U(2) flavor symmetry
		\\*[20pt]}
	}

	\author{
  Morimitsu Tanimoto $^{1}$
  \footnote{email: tanimoto@muse.sc.niigata-u.ac.jp}
      \hskip 0.5 cm and \hskip 0.5 cm
       Kei Yamamoto $^{2}$
   \footnote{e-mail: k.yamamoto.e3@cc.it-hiroshima.ac.jp}
	\\*[20pt]
	\centerline{
	\begin{minipage}{\linewidth}
	\begin{center}
	$^1${\it \normalsize
	Department of Physics, Niigata University, Niigata 950-2181, Japan }
	\\*[10pt]
				$^2${\it \normalsize  Department of Global Environment Studies, Hiroshima Institute of Technology, Hiroshima
					731-5193, Japan} 
	\end{center}
	\end{minipage}}
	\\*[30pt]}
		\date{\today\\	\vskip 3 cm		
		{\small \bf Abstract}
		\begin{minipage}{0.9\linewidth}
			\medskip
			\medskip
			\small
		We study the interplay of  New Physics (NP) among
	the  lepton  magnetic moment, the  lepton flavor violation (LFV)  and the electron electric dipole moment (EDM)   in light of recent data of the muon $(g-2)_\mu$. The NP is discussed in the leptonic dipole operator
	with the $U(2)$ flavor symmetry of the charged leptons,
	where  possible CP violating phases of the  three family space are taken into account.
	It is remarked that  the third-family 
	contributes significantly to  the LFV decay,
	$\mu \to e\gamma$,	and the electron EDM.
	The experimental upper-bound on $\mu \to e\gamma$ decay gives 
	a severe constraint on the parameters of the flavor model.	
	The predicted electron EDM is rather large	due to  the CP violating phases in the three family space.
	In addition, we also study  $(g-2)_{e,\tau}$
	of the  electron and tauon, and EDMs of the muon and  tauon as well as the $\tau \to  e \gamma$ and  $\tau \to \mu \gamma$
	decays.
	The  $\tau_R \to \mu_L \gamma$ decay is predicted 
	 to be close to the experimental upper-bound. 
	\end{minipage}}
	
	\begin{titlepage}
		\maketitle
		\thispagestyle{empty}
	\end{titlepage}


\section{Introduction}

The electric and  magnetic dipole moments  of 
the electron and the muon are  low-energy probes 
of New Physics (NP) beyond the Standard Model (SM). 
Recently, the muon $(g-2)_\mu$ experiment at Fermilab reported  a new measurement of the muon magnetic anomaly using data collected in 2019 (Run-2) and 2020
(Run-3) \cite{Muong-2:2023cdq}.  
The improved analysis and run condition lead to more than a factor of two reduction in the systematic
uncertainties, which is compared to 
 the E989 experiment at Fermilab~\cite{Muong-2:2021ojo} and the previous BNL result~\cite{Muong-2:2006rrc}. 
This result indicates a $5.1\,\sigma$ discrepancy with the SM prediction
by "Muon $g-2$ Theory Initiative" \cite{Aoyama:2020ynm}
(see also \cite{Jegerlehner:2017gek,Colangelo:2018mtw,Hoferichter:2019mqg,Davier:2019can,Keshavarzi:2019abf,Hoid:2020xjs,Czarnecki:2002nt, Melnikov:2003xd,Aoyama:2012wk, Gnendiger:2013pva}).

However, there is a debatable point 
	on the precise value of the SM prediction:
	the problem is  the contribution of the hadronic vacuum polarization (HVP). 
The current situation is complicated.
The CMD-3 collaboration \cite{CMD-3:2023alj}  released results on the 
cross section that disagree at the $(2.5-5) \sigma$ level with all previous measurements, including those from the previous  CMD-2 collaboration.
 The origin of this discrepancy is currently unknown.
The BMW collaboration published the first complete lattice-QCD result with subpercent precision  \cite{Borsanyi:2020mff}.
Their result is closer to the experimental average
of the muon $(g-2)_\mu$ measurements, showing only a  $1.7\sigma$ 
preference 
for  NP in the muon $(g-2)_\mu$.
 Further studies are underway to clarify these theoretical differences. 
 The white paper (the SM prediction) is expected to be updated 
 including the HVP problem  in 2024.


If the muon $(g-2)_\mu$ anomaly is due to NP, its effect could appear 
in other observables of the charged lepton sector. 
The interesting one is the electric dipole moments (EDM)  of the electron. 
This year, the JILA group has reported a new upper-bound on the electron EDM, which is $|d_e|<4.1 \times 10^{-30} \rm e\, cm \,(90\%\, { confidence})$,
by using the  $\rm  Hf\, F^+$ ions
	trapped by the rotating electric field \cite{Roussy:2022cmp}.
	It overcame the latest ACME collaboration result obtained in 2018
 \cite{Andreev:2018ayy}.
	 Precise measurements of  the electron EDM will be rapidly updated in the future. The future sensitivity
	  at ACME $\rm I\hskip -0.04cm I\hskip -0.04cm I$ is expected to be
	  $|d_e|<0.3\times10^{-30}\,\text{e\,cm}$
	 \cite{Kara:2012ay,ACMEIII}.
	 In contrast, the present upper-bound of the muon \cite{Muong-2:2008ebm} and tauon EDM's \cite{Belle:2002nla,Bernreuther:2021elu,Uno:2022xau} are not so tight.	
 
  The lepton flavor violation (LFV) is also  possible NP phenomena of 
   the charged leptons.
  The most severe constraint on LFV is the branching ratio of the $\mu\to e\gamma$ decay.
  The experimental  bound is $\mathcal{B}({\mu^+ \to e^+ \gamma}) < 4.2 \times 10^{-13}$ from the MEG experiment~\cite{TheMEG:2016wtm}.
On the other hand, the current upper-bounds for 
 $\mathcal{B}\,({\tau \to \mu \gamma})$
 and $\mathcal{B}\,({\tau \to e \gamma})$ are 
  $4.4~ \times~ 10^{-8}$  and $3.3~ \times~ 10^{-8}$ \cite{BaBar:2009hkt,Belle:2021ysv},
respectively.

Comprehensive  studies 
of the electric and  magnetic dipole moments  of leptons 
are given in the SM Effective Field Theory (SMEFT)
\cite{Buchmuller:1985jz,Grzadkowski:2010es,Alonso:2013hga} under the hypothesis of new degrees of freedom
above the electroweak scale \cite{Panico:2018hal,Aebischer:2021uvt,Allwicher:2021rtd,Kley:2021yhn}. 
The phenomenological discussion of NP has presented taking into account  the muon $(g-2)_\mu$ anomaly and the LFV bound in the SMEFT 
imposing  {$U(2)_{L_L} \otimes U(2)_{E_R}$}  flavor symmetry \cite{Isidori:2021gqe},
which is the maximally allowed  subgroup of
 $U(3)_{L_L} \otimes U(3)_{E_R}$
 \footnote{The $U(3)$ flavor symmetry is the maximal flavor symmetry allowed by the SM gauge sector.
 It allows us to apply the Minimal Flavor Violation hypothesis~\cite{Chivukula:1987py,DAmbrosio:2002vsn}.
  } 
  acting only on the first two (light) families \cite{Barbieri:2011ci,Barbieri:2012uh,Blankenburg:2012nx}. 
Since this  flavor symmetry  reduces the number of independent parameters of the flavor sector \cite{Fuentes-Martin:2019mun,Faroughy:2020ina}, we can expect the interplay among possible NP evidence in the observables of different flavors,
 which is independent of  the details of NP.
Thus,  the SMEFT with flavor symmetry  can probe NP without discussing its dynamics in the flavor space.

We have already studied the interplay of NP
among  the muon $(g-2)_\mu$, the electron EDM and  the  $\mu\to e\gamma$ decay in the modular symmetry of flavors \cite{Kobayashi:2021pav,Kobayashi:2022jvy},
 which is realized in the context of string effective field theory \cite{Kobayashi:2021uam} (see also \cite{Calibbi:2021qto}).
The results obtained are dependend on 
 the modular weights of the charged leptons and the relevant discrete symmetry,
  considerably.
 Therefore, more general discussions are required to confirm the role of flavor symmetry in the NP search
 without depending on the details of the flavor model.

 There are many possible discrete symmetries  for the leptons flavor mixing
 \cite{Altarelli:2010gt,Ishimori:2010au,Ishimori:2012zz,Kobayashi:2022moq,Hernandez:2012ra,King:2013eh,King:2014nza,Tanimoto:2015nfa,King:2017guk,Petcov:2017ggy,Feruglio:2019ybq}.
 Some of them are subgroups of continuous groups 
 \cite{Kobayashi:2022moq}.
 On the other hand, NP analysis of the SMEFT has been  attractively performed 
  by imposing the flavor symmetry of continuous group, especially  $U(2)$
  \cite{Fuentes-Martin:2019mun,Faroughy:2020ina}.
We also  adopt  the $U(2)$ flavor symmetry for 
discussing the dipole operator of leptons
 to investigate NP \cite{Isidori:2021gqe}.
In this paper, we present numerical discussions of the interplay of  NP  
in light of recent data of the muon $(g-2)_\mu$ and the electron EDM
by taking   {$U(2)_{L_L} \otimes U(2)_{E_R}$} flavor symmetry 
of the charged leptons. 
Since we  discuss the electron EDM,
  we analyze  NP in the three family space  of
   {$U(2)_{L_L} \otimes U(2)_{E_R}$} flavor symmetry, taking into account the possible CP violating phases.
  It is remarked that  the third-family
   contributes significantly to the $\mu \to e\gamma$ decay and  the electron EDM.
In addition to the muon $(g-2)_\mu$ and the electron EDM,
we also study  $(g-2)_{e,\tau}$ of  the  electron and tauon,
 and EDMs of the muon and  tauon as well as LFV processes
$\tau \to  e \gamma$ and  $\tau \to \mu \gamma$.

The paper is organized as follows.
In section \ref{sec:data constraints},
	we discuss  the experimental constraints on the Wilson coefficients of 
	the leptonic dipole operator. 
	Section \ref{U2 model} presents our framework of the $U(2)$ 
	flavor model. 
	Section \ref{numerical} provides numerical discussions.
		The summary and discussion are devoted to section \ref{summary}.
		In  Appendix \ref{appen:exp}, the experimental constraints
	of the relevant Wilson coefficients of the leptonic  dipole operator
	are presented.
	In Appendix \ref{YLLRR}, the  $3\times 3$ Yukawa matrices,
	 $YY^\dagger$ and $Y^\dagger Y$, are given	explicitly.

\section{Constraints of Wilson coefficients of the  dipole operator}
\label{sec:data constraints}

\subsection{Input experimental data}

The combined result  
    from the E989 experiment at Fermilab ~\cite{Muong-2:2023cdq,Muong-2:2021ojo} and the E821 experiment at BNL~\cite{Muong-2:2006rrc} on $a_\mu=(g-2)_\mu/2$,
together with the SM prediction $a_\mu^\mathrm{SM}$ in~\cite{Aoyama:2020ynm}, implies
\begin{align}
\Delta a_\mu = a_\mu^\mathrm{Exp} - a_\mu^\mathrm{SM} = \brackets{249 \pm 49} \times 10^{-11}~.
\label{muon-data}
\end{align}
We suppose that the contribution of NP appears in $\Delta a_\mu$.

However,
the precise value of the SM prediction of HVP is still  unclear.
Further studies will clarify the theoretical differences. 
If $\Delta a_\mu$ is significantly lower, less than or of order $1\,\sigma$,
one may obtain somewhat loose bound in phenomenological studies of NP \cite{Shafi:2023ksr}.
In our study, we take a following reference  value
(the discrepancy of  $5.1\,\sigma$ with the SM prediction)  as the input in our numerical analysis:
\begin{align}
\Delta a_\mu &=  249 \times 10^{-11}\,.
\label{muon-input}
\end{align}
We will comment on our result in the case that the discrepancy with the SM prediction is reduced  in section \ref{summary}.
 
We also input the upper-bound  of the electron EDM
by the JILA group \cite{Roussy:2022cmp}:
\begin{align}
|{d_e} |<4.1 \times 10^{-30} \, \rm  e\,cm =6.3\times 10^{-14} \,TeV^{-1}\,.
\label{eEDM-input}
\end{align} 
On the other hand, 
the upper-bound of the muon EDM
is \cite{Muong-2:2008ebm}:
\begin{align}
|{d_\mu} |<1.8 \times 10^{-19} \, \rm  e\,cm =2.76\times 10^{-3} \,TeV^{-1}\,.
\label{muEDM}
\end{align} 
 
 The tauon EDM can be evaluated through the measurement of CP-violating correlations in tauon-pair production such as $e^+e^-\to \tau^+\tau^-$
 \cite{Belle:2002nla} (see also \cite{Bernreuther:2021elu}).
 The present upper-bound on the tauon EDM $d_\tau$ is given as: \cite{Uno:2022xau}:
 \begin{align}
-1.85\times  10^{-17}{\rm e\,cm}<{\rm Re} \,d_\tau<
0.61\times  10^{-17}{\rm e\,cm},\ \
-1.03\times  10^{-17}{\rm e\,cm}<{\rm Im}\, d_\tau<
0.23\times  10^{-17}{\rm e\,cm}.
 \label{tauEDM0}
 \end{align} 
 Taking the bound of ${\rm Re} \,d_\tau$, we have 
  \begin{align}
 |d_\tau|<1.85\times  10^{-17}{\rm e\,cm}=2.84\,{\rm TeV}^{-1}\,.
 \label{tauEDM}
 \end{align}

The experimental upper-bound for the branching ratio of $\mu\to e\gamma$ is \cite{TheMEG:2016wtm}:
\begin{align}
\mathcal{B}({\mu^+ \to e^+ \gamma}) < 4.2 \times 10^{-13}\,.
\label{LFV-input}
\end{align} 
We also take account of the upper-bound for LFV comes from the branching ratios
of $\tau\to\mu\gamma$  and  $\tau\to e\gamma$ \cite{BaBar:2009hkt,Belle:2021ysv} :
\begin{align}
\mathcal{B}({\tau \to \mu \gamma}) < 4.2 \times 10^{-8}\,,\quad\qquad
\mathcal{B}({\tau \to e \gamma}) < 3.3 \times 10^{-8}\,.
\label{LFV-input2}
\end{align} 

These input data are converted into the magnitudes of the Wilson coefficients of the  leptonic dipole operator in the next subsection.

\subsection{Wilson coefficients of the  leptonic dipole operator}
We make the assumption that NP is heavy and can be given by the
SMEFT Lagrangian.
Let us focus on  the dipole operators of leptons and their Wilson coefficients  at the weak scale as:
\begin{align}
&\mathcal{O}_{\underset{LR}{e\gamma}}
= \frac{v }{\sqrt{2}}  \overline{E}_{L}  \sigma^{\mu\nu} E_{R} F_{\mu\nu}\,,\qquad\qquad
\CL{e\gamma}{LR}^\prime=
\begin{pmatrix}
\CL{e\gamma}{ee}^\prime &\CL{e\gamma}{e\mu}^\prime
&\CL{e\gamma}{e\tau}^\prime\\
\CL{e\gamma}{\mu e}^\prime &\CL{e\gamma}{\mu\mu}^\prime
&\CL{e\gamma}{\mu\tau}^\prime\\
\CL{e\gamma}{\tau e}^\prime &\CL{e\gamma}{\tau\mu}^\prime
&\CL{e\gamma}{\tau\tau}^\prime\\
\end{pmatrix} 
\,,
\nonumber\\
&
\mathcal{O}_{\underset{RL}{e\gamma}}
= \frac{v }{\sqrt{2}}  \overline{E}_{R}  \sigma^{\mu\nu} E_{L} F_{\mu\nu}\,,\qquad\qquad
\CL{e\gamma}{RL}^\prime=\CL{e\gamma}{LR}^{\prime\, \dagger}
\,,
\label{dipole-operators}
\end{align}
where $E_L$ and $E_R$ denote three flavors of the left-handed and
right-handed leptons, respectively,
and $v$ denotes the vacuum expectation value (VEV) of the Higgs field $H$.
The prime of the Wilson coefficient indicates the {flavor} basis corresponding to the mass-eigenstate basis of the charged leptons.
The relevant effective Lagrangian is written as:
\begin{align}
\mathcal{L}_{\rm dipole}=
\frac{1}{\Lambda^2}\,\left (
\CL{e\gamma}{LR}^\prime\mathcal{O}_{\underset{LR}{e\gamma}}
+\CL{e\gamma}{RL}^\prime\mathcal{O}_{\underset{RL}{e\gamma}}
\right )
\,,
\end{align}
where $\Lambda$ is a certain mass scale of NP  in the effective theory.
Here the Wilson coefficient is understood to be evaluated at the weak scale 
(we neglect the small effect of running below the weak scale)
 \footnote{The one-loop effect is small as seen in~\cite{Buttazzo:2020ibd}.}.

Inputting the value  in Eq.\,\eqref{muon-input},
the Wilson coefficient is obtained as\cite{Isidori:2021gqe} 
(see  Appendix \ref{appen:exp}):
\begin{align}
\frac{1}{\Lambda^2}\text{Re}\  [\CL{e\gamma}{\mu\mu}^\prime]
= 1.0 \times 10^{-5} \, \mathrm{TeV}^{-2} \,.
\label{Cmumu-exp}
\end{align}

The LFV process  $\mu \to e \gamma$ gives us more severe constraint
for the Wilson coefficient by the experimental data in Eq.\,\eqref{LFV-input}.
The upper-bound is obtained \cite{Isidori:2021gqe} 
(see Appendix \ref{appen:exp}):
\begin{align}
	\frac{1}{\Lambda^2}|\CL{e\gamma}{e\mu(\mu e)}^\prime| <  2.1 \times 10^{-10} \, \mathrm{TeV}^{-2} \, .
	\label{eq:bound_C_egamma_12}
\end{align}
Taking into account Eqs.~(\ref{Cmumu-exp}) and \eqref{eq:bound_C_egamma_12}, one has the ratio \cite{Isidori:2021gqe} :
\begin{align}
\left |\frac{ \CL{e\gamma}{e\mu(\mu e)}^\prime  }{    \CL{e\gamma}{\mu\mu}^\prime    }\right | <  2.1\times 10^{-5}\,.
\label{eq:bound12}
\end{align}
Thus, the magnitude of  $\CL{e\gamma}{e\mu(\mu e)}^\prime $
is much suppressed compared with $\CL{e\gamma}{\mu\mu}^\prime $.
This  gives the severe constraint for parameters of the flavor  model.

The   $\tau \to \mu \gamma$ process also  gives us another  constraint
for the Wilson coefficient by the experimental data in Eq.\,\eqref{LFV-input2}.
The upper-bounds are obtained as seen in Appendix \ref{appen:exp}:
\begin{align}
\frac{1}{\Lambda^2}|\CL{e\gamma}{\mu\tau(\tau\mu)}^\prime| <  2.65 \times 10^{-6} \, \mathrm{TeV}^{-2} \,, \qquad 
\frac{1}{\Lambda^2}|\CL{e\gamma}{e\tau(\tau e)}^\prime| <  2.35 \times 10^{-6} \, \mathrm{TeV}^{-2}  .
\label{eq:bound_C_egamma_23-13}
\end{align}

The electron EDM,  $d_e$ is defined in the operator:
\begin{align}
\mathcal{O}_{\mathrm{edm}}
= -\frac{i}{2}\,d_e(\mu) \, \overline{e}  \sigma^{\mu\nu} \gamma_5 e F_{\mu\nu}\,,
\label{EDM}
\end{align}
where $d_e=d_e(\mu=m_e)$.
Therefore, the electron EDM is extracted from the effective Lagrangian 
\begin{align}
\mathcal{L}_{\rm EDM}=
\frac{1}{\Lambda^2}\CL{e\gamma}{ee}^\prime \mathcal{O}_{\underset{LR}{e\gamma}} + \rm h.c.
= \frac{1}{\Lambda^2}\CL{e\gamma}{ee}^\prime \frac{v }{\sqrt{2}}  \overline{e}_{L}  \sigma^{\mu\nu} e_R F_{\mu\nu}
+ \rm h.c.\,,
\label{SMEFT-EDM0}
\end{align}
which leads to 
\begin{align}
d_e=-\sqrt{2} \, \frac{v}{\Lambda^2}\, {\rm Im}\, [\CL{e\gamma}{ee}^\prime] \,,
\label{SMEFT-EDM}
\end{align}
at tree level, where the small effect of running below the electroweak scale is neglected.

Inputting the experimental  upper-bound of the electron EDM
in Eq.\,\eqref{eEDM-input} \cite{Roussy:2022cmp},
we obtain the constraint of the Wilson coefficient:
\begin{equation}
	\frac{1}{\Lambda^2}{\mathrm{Im}}\, [\CL{e\gamma}{ee}^\prime]<1.8\times 10^{-13}\, 
	\mathrm {TeV^{-2}}\,.
	\label{Cee-exp}
\end{equation}
On the other hand,  the experimental  upper-bound of the muon EDM
in Eq.\,\eqref{muEDM}  gives:
\begin{equation}
\frac{1}{\Lambda^2}{\mathrm{Im}}\, [\CL{e\gamma}{\mu \mu}^\prime]<7.9\times 10^{-3}\, 
\mathrm {TeV^{-2}}\,.
\end{equation}
\begin{table}[t!]
	\begin{center}
		\renewcommand{\arraystretch}{1.1}
		\begin{tabular}{c|l|l} 
			Observables &  Exp.$-$ SM / upper bound & Wilson Coef. in $1/\Lambda^2\ [\mathrm{TeV}^{-2}]$  \\ \hline
			$\Delta a_\mu$ & $249 \times 10^{-11}$\cite{Muong-2:2023cdq,Muong-2:2021ojo,Muong-2:2006rrc} 
			& $\text{Re}\  [\CL{e\gamma}{\mu\mu}^\prime] =1.0 \times 10^{-5} 
			$\\ 
			$\mathcal{B}({\mu^+ \to e^+ \gamma}) $& $< 4.2 \times 10^{-13}$ \cite{TheMEG:2016wtm}
			& $|\CL{e\gamma}{e\mu(\mu e)}^\prime| <2.1 \times 10^{-10}  $\\
			$\mathcal{B}({\tau \to \mu \gamma}) $& $< 4.2 \times 10^{-8}$ \cite{BaBar:2009hkt,Belle:2021ysv}
			& $ | \CL{e\gamma}{\mu \tau(\tau\mu)}^\prime| <2.65 \times 10^{-6}  $\\
			$\mathcal{B}({\tau \to e \gamma}) $&  $< 3.3 \times 10^{-8}$ \cite{BaBar:2009hkt,Belle:2021ysv}
			& $|\CL{e\gamma}{e\tau(\tau e)}^\prime|<  2.35 \times 10^{-6}$ \\ 
			$|d_e| $ & $<4.1 \times 10^{-30} \, \rm  e\,cm$ \cite{Roussy:2022cmp}&$\text{Im}\  [\CL{e\gamma}{ee}^\prime]<1.8\times 10^{-13}$ \\ 
			$|d_\mu| $ & $<1.80 \times 10^{-19} \, \rm  e\,cm$\cite{Muong-2:2008ebm}  &$\text{Im}\  [\CL{e\gamma}{\mu\mu}^\prime]<7.9\times 10^{-3} $ \\ 
			$|d_\tau |$ & $<1.85 \times 10^{-17} \, \rm  e\,cm$ \cite{Belle:2002nla}  &$\text{Im}\  [\CL{e\gamma}{\tau\tau}^\prime]<8.2\times 10^{-1} $ 
		\end{tabular}
	\end{center}
	\caption{Relevant observables and the corresponding values of Wilson coefficients, which
		are presented in $1/\Lambda^2\  (\mathrm{TeV}^{-2})$ unit.}
	\label{tab:data}
\end{table}
The  upper-bound of the tauon EDM
in Eq.\,\eqref{tauEDM} also gives:
\begin{equation}
\frac{1}{\Lambda^2}{\mathrm{Im}}\, [\CL{e\gamma}{\tau \tau}^\prime]<8.2\times 10^{-1}\,  \mathrm {TeV^{-2}}\,.
\end{equation}
Using these bounds of Wilson coefficients, we analyze the $U(2)$ flavor model in the next section.
These data are listed  in Table \ref{tab:data}.

\section{U(2) flavor model }
\label{U2 model}
\subsection{Flavor structure of Yukawa and dipole operator in  charged leptons}
In the lepton sector,
 we impose {$U(2)_{L_L} \otimes U(2)_{E_R}$} flavor symmetry,
which is the corresponding  subgroup acting only on the first two light families.
The minimal set of the breaking terms, which are 
so called spurions, are
\footnote{See  the discussion to get this minimal set
	of spurions in the section 2 of  \cite{Barbieri:2011ci}.
}
\begin{align}
V_\ell\sim ({\bf 2,\,1})\,,\qquad \Delta_e \sim ({\bf 2,\,\bar 2})\,.
\end{align}
The two suprions can be parameterized by
\begin{align}
V_\ell=
\begin{pmatrix}
0\\
{\epsilon_{\ell}}\,
\end{pmatrix}
,\qquad \Delta_e=O_e^T \ 
\begin{pmatrix}
\delta'_e & 0\\
0& \delta_e\\
\end{pmatrix}\,,\qquad 
O_e=
\begin{pmatrix}
c_e & s_e\\
-s_e& c_e\\
\end{pmatrix}\,,
\label{spurions}
\end{align}
where $\delta'_e$,  $\delta_e$ and  $\epsilon_\ell$ are taken to be real
 without loss of generality,
 while $c_e$ and $s_e$ denote $\cos \theta_e$ and  $\sin \theta_e$,
 respectively.
 
 In this scheme, the flavor structure of the dipole operator has already discussed in Ref.\cite{Isidori:2021gqe}.
  	The analysis was done in two family space with 1st-2nd- and 2nd-3rd-sector, separately.
	  However, as will be discussed later, in the three family space, 
	 the third generation gives significant effects on the 1st-2nd sector.
	Discussing  the flavor structure in three family space,
we can present  the systematic discussion
 of the interplay among the lepton  $(g-2)_\ell$, EDMs and  LFV decays.
 Especially, we can also discuss the EDM of the electron
  due to non-trivial CP phases.

 
Using the spurions in Eq.\,\eqref{spurions}, the $3\times 3$ Yukawa coupling of  leptons $Y_e$ is written
in the 
left-right (LR) convention at order ${\cal O}(V_\ell^2 \Delta_e)$ of  spurion couplings:
%
 \begin{align}
 Y_e= Y_{e0}
 \begin{pmatrix}
 C^y_{\Delta} (\Delta_e)_{\alpha\beta} + C^y_{VV\Delta}  (V_\ell)_\alpha(V_\ell^\dagger)_\gamma (\Delta_e)_{\gamma\beta}
 & C^y_{V} (V_\ell)_\alpha
  \\
 C^y_{V\Delta} (V_\ell^\dagger)_\alpha (\Delta_e)_{\alpha\beta} & C^y
 \end{pmatrix}_{LR} \, ,
 \label{Y-coupling}
  \end{align}
  where $\alpha,\beta$ are the first or second component in Eq.\,\eqref{spurions}.
 Next-to-leading-order spurion couplings as $C^y_{VVV}V_\ell V_\ell^\dagger V_\ell$ and $C^y_{VV} V_\ell^\dagger V_\ell$ can be added, but are not included here because they do not add any independent structure.
 The coefficients  $C^y$, $C^y_{\Delta}$, $C^y_{V}$, $C^y_{V\Delta}$
 and $C^y_{VV\Delta}$ 
  are complex parameters of order $1$,
 and $Y_{e0}$ is a normalization factor to realize the observed tauon mass.
 Taking spurion parameters in Eq.\,\eqref{spurions}, the Yukawa matrix
  is written explicitly as:
  \begin{align}	
 Y_e=  Y_{e0} \begin{pmatrix}
 C^y_{\Delta} c_e \delta_e^\prime & -C^y_{\Delta} s_e \delta_e & 0 \\
 s_e \delta_e^\prime (C^y_{\Delta}+C^n_{VV\Delta}\epsilon_\ell^2) & c_e \delta_e (C^y_{\Delta}+C^y_{VV\Delta}\epsilon_\ell^2) 
 & C^y_{V} \epsilon_\ell
 \\
 C^y_{V\Delta} (s_e \epsilon_\ell \delta_e^\prime) & C^y_{V\Delta} (c_e \epsilon_\ell \delta_e) 
 & C^y
 \end{pmatrix}_{LR} \,.
 \label{U2y}
 \end{align}

The flavor structure of the leptonic  dipole operator
in Eq.\,\eqref{dipole-operators} is given by $3\times 3$ matrix $X^{e\gamma}$.
It is  also given in terms of the spurion couplings like
in Eq.\,\eqref{Y-coupling}
as follows:

\begin{align}
X^{e\gamma}  &= 
 \begin{pmatrix}
		C^{e\gamma}_{\Delta} (\Delta_e)_{\alpha\beta} + C^{e\gamma}_{VV\Delta} (V_\ell)_\alpha (V_\ell^\dagger)_\gamma (\Delta_e)_{\gamma\beta}
		&C^{e\gamma}_{V} (V_\ell)_\alpha
		\nonumber\\
		C^{e\gamma}_{V\Delta} (V_\ell^\dagger)_\alpha (\Delta_e)_{\alpha\beta}& C^{e\gamma}
	\end{pmatrix}_{LR} \, \nonumber\\
	&= \begin{pmatrix}
		C^{e\gamma}_{\Delta} c_e \delta_e^\prime & -C^{e\gamma}_{\Delta} s_e \delta_e & 0 \\
		s_e \delta_e^\prime (C^{e\gamma}_{\Delta}+C^{e\gamma}_{VV\Delta}\epsilon_\ell^2) & c_e \delta_e (C^{e\gamma}_{\Delta}+C^{e\gamma}_{VV\Delta}\epsilon_\ell^2) 
		& C^{e\gamma}_{V} \epsilon_\ell 
		 \\
		C^{e\gamma}_{V\Delta} (s_e \epsilon_\ell \delta_e^\prime)
	 & C^{e\gamma}_{V\Delta} (c_e \epsilon_\ell \delta_e) 
		& C^{e\gamma}
	\end{pmatrix}_{LR} \,,
	\label{U2egamma}
\end{align}
where $C^{e\gamma}$, $C^{e\gamma}_{\Delta}$, $C^{e\gamma}_{V}$, $C^{e\gamma}_{V\Delta}$ and
$C^{e\gamma}_{VV\Delta}$
are 
also complex parameters of order $1$.
Since the Wilson coefficients in Eq.\,\eqref{dipole-operators}
are  written in  the mass-eigenstate basis of the charged leptons,
 the matrix $X^{e\gamma}$ in Eq.\eqref{U2egamma}
 should be transformed into the diagonal basis of
 $Y$ in Eq.\eqref{U2y}.

The eigenvalues of Yukawa matrix $Y$ in Eq.\eqref{U2y}
($y_{e}\ll y_{\mu}\ll y_{\tau}$) are obtained by
solving the eigenvalue equation.
For the determinant and trace of $ Y Y^\dagger $,  
one finds in the leading order:
\begin{align}
&{\rm Det}\,[Y Y^\dagger ]=y^2_{e}\, y^2_{\mu}\, y^2_{\tau} \simeq
	Y_{e0}^6\,|C^{y}_{\Delta}|^4 \,|C^y|^2\, c_e^4\,  \delta_e^2\,  { \delta'_e}^2 \,,
	\nonumber\\
&{\rm Tr}\,[Y Y^\dagger ]=y^2_{e}+y^2_{\mu}+y^2_{\tau} \simeq Y_{e0}^2\, |C^y|^2\,,	
\label{eq:DetTrace}
\end{align}
where $\delta_e$ and $\delta'_e$
 are much smaller than $1$ to reproduce the charged lepton mass hierarchy.
An expression for $y_{\mu}^2 y_{\tau}^2$ is given by the determinant of the
2-3 submatrix:
\begin{align}
y_{\mu}^2\, y_{\tau}^2\simeq Y_{e0}^4\,|C^{y}_{\Delta}|^2 \,|C^y|^2\, c_e^2\,  \delta_e^2\,.
\label{subdet}
\end{align}
%
Then, one gets
\begin{align}
y_{\tau}^2\simeq Y_{e0}^2\,|C^y|^2 \,,\qquad 
y_{\mu}^2\simeq Y_{e0}^2\,|C^{y}_{\Delta}|^2  c_e^2\,  \delta_e^2\,,\qquad
y_{e}^2\simeq Y_{e0}^2\,|C^{y}_{\Delta}|^2  c_e^2\, { \delta'_e}^2\,,
\qquad
\label{eigenvalues}
\end{align}
which lead to
\begin{align}
\frac{y_{e}^2}{y_{\mu}^2}\simeq \frac{ { \delta'_e}^2}{ \delta_e^2}\,,
\qquad 
\frac{y_{\mu}^2}{y_{\tau}^2}
\simeq \frac{  |C^{y}_{\Delta}|^2}{|C^{y}|^2}\,c_e^2\delta_e^2
\simeq  \delta_e^2\,.
\label{massratio}
\end{align}
These ratios indicate $ \delta_e\gg   \delta'_e$
to reproduce the Yukawa hierarchy of the charged leptons.
Especially, ${y_{e}^2}/{y_{\mu}^2}$ is  independent of other
  coefficients of order one.

While the magnitudes of $\delta_e$ and $ \delta'_e$  can be constrained  by the charged lepton Yukawa couplings,
$s_e$ and $\epsilon_\ell$ cannot be determined directly
\footnote{These parameters of the $U(2)$ flavor model 
		were discussed in the view of NP in the semileptonic $B$-physics anomalies \cite{Bordone:2017anc}. 
}.
Since these parameters are the  $U(2)$ breaking ones,
the most natural choice is \cite{Blankenburg:2012nx,Faroughy:2020ina},  
\begin{align}
s_e\sim s_q\sim 0.01-0.1\,,  
\qquad  \epsilon_\ell \sim \epsilon_q \sim 0.01-0.1\,,
\label{s-ep}
\end{align}
in the similar treatment of quark and lepton sectors,
where $s_q$ and  $\epsilon_q$ denote parameters of spurions of
the  quark sector.  

The neutrino mass matrix was  given in $U(2)$  flavor model
to reproduce observed large mixing angles of leptons
in view  of quark-lepton unification 
\cite{Carone:1997qg,Tanimoto:1997zw,Blazek:1999ue}.
Also the large mixing angles have been discussed by introducing relevant spurions in $U(2)$ and $U(3)$ flavor model \cite{Blankenburg:2012nx}.
 However, we do not address details of neutrino mass matrix in this work
because its contribution to our result is negligibly small due to small neutrino masses.

\subsection{Mass-eigenstate basis of the charged leptons}

The Yukawa matrix $Y$ in Eq.\,\eqref{U2y}
 is diagonalized by the unitary transformation
 $U_L^\dagger Y U_R$, where
 the unitary matrices are given in terms of
 $2\times 2$ orthogonal matrices   and phase matrices:
 \begin{align}
 U_L=P_{L23} U_{L23} P_{L13} U_{L13} P_{L12} U_{L12}\,,\qquad
 U_R=P_{R23} U_{R23} P_{R13} U_{R13} P_{R12} U_{R12}\, .
 \label{unitary}
 \end{align}
The rotation matrices are 
  \begin{align}
 U_{12}=
 \begin{pmatrix}
 c_{12} & s_{12} &0\\
 -s_{12}& c_{12}&0\\
 0&0&1\\
 \end{pmatrix}\,, \qquad
  U_{13}=
 \begin{pmatrix}
 c_{13} &0& s_{13} \\
  0&1&0\\
 -s_{13} &0& c_{13} \\
 \end{pmatrix}\,,\qquad
  U_{23}=
 \begin{pmatrix}
 1 &0& 0 \\
 0&c_{23} & s_{23} \\
 0&-s_{23} & c_{23} \\
 \end{pmatrix}\,,
 \end{align}
 and phase matrices are 
  \begin{align}
 P_{12}=
 \begin{pmatrix}
 1 & 0 &0\\
 0& e^{i \phi_{12}}&0\\
 0&0&1\\
 \end{pmatrix}\,, \qquad
 P_{13}=
 \begin{pmatrix}
 1 & 0 &0\\
 0& 1&0\\
 0&0&e^{i \phi_{13}}\\
 \end{pmatrix}\,, \qquad
 P_{23}=
\begin{pmatrix}
1 & 0 &0\\
0& 1&0\\
0&0&e^{i \phi_{23}}\\
\end{pmatrix}\, ,
 \end{align}
 where  notations $L$ and $R$ are omitted.
  The $s_{ij}$ and $\phi_{ij}$ are obtained approximately
  from $Y Y^\dagger$ or $Y^\dagger Y$ in Appendix \ref{YLLRR}
  for the left-handed or the right-handed sector.
  
  In the following expression,
   we take $C^y_\Delta$ to be  real positive and  $C^y$, $C^y_{V}$,  $C^y_{V \Delta}$ and   $C^y_{VV\Delta}$  to be  complex without loss of generality.
  The mixing angles $s_{Lij}$ and $s_{Rij}$ are  obtained approximately  
   (see Appendix \ref{YLLRR}). The mixing angles  $s_{L12}$ and $s_{R12}$ are :
 \begin{align}
 &s_{L12}\simeq -\frac{s_e}{c_e} \left [1-\left| \frac{C^y_{VV\Delta}}{C^y_\Delta}\right |\epsilon_\ell^2 
 \cos (\arg C^y_{VV\Delta})  \right]\,, \nonumber\\
 &s_{R12}\simeq 2\frac{s_e}{c_e} \frac{\delta'_e}{\delta_e}
 \left [\cos (\arg C^y_{VV\Delta}) +
 \left| \frac{C^y_{VV\Delta}}{2C^y_\Delta}\right |\epsilon_\ell^2   \right]
  \left| \frac{C^y_{VV\Delta}}{C^y_\Delta}\right |\epsilon_\ell^2  \,,
 \end{align}
respectively.
The phase  $\phi_{L12}$ and $\phi_{R12}$ are:
 \begin{align}
\phi_{L12}\simeq  -\left| \frac{C^y_{VV\Delta}}{C^y_\Delta}\right |\epsilon_\ell^2 
\sin (\arg C^y_{VV\Delta}) \,, \qquad
\phi_{R12}\simeq 0\,,
\label{U12}
\end{align}
which are much smaller than $1$.
The mixing angles  $s_{L23}$ and $s_{R23}$ are rather simple
in the leading order  as:
\begin{align}
s_{L23}\simeq \left | \frac{C^y_{V}}{C^y} \right | \epsilon_\ell \,, \qquad
s_{R23}\simeq \left|\frac{C^{y\,*}_\Delta C^y_V+C^{y\,*}_{V\Delta}C^y}{C^{y\, 2}}
 \right |c_e \delta_e \epsilon_\ell  \,,
\end{align}
respectively.
The phase  $\phi_{L12}$ and $\phi_{R12}$ are:
\begin{align}
\phi_{L23}\simeq \arg \frac{C^y_{V}}{C^y} \,, \qquad
\phi_{R23}\simeq 
\arg \left [\frac{C^{y\,*}_\Delta C^y_V+C^{y\,*}_{V\Delta}C^y}{|C^{y}|^2}
\right ] \,,
\label{U23}
\end{align}
which are of order $1$.

The mixing angles  $s_{L13}$ and $s_{R13}$ are also simple
in the leading order  as:
\begin{align}
s_{L13}\simeq \left | \frac{C^y_{\Delta}C^{y\,*}_{V\Delta}}{C^{y\,2}} \right 
| c_e s_e (\delta_e^2 -{\delta'_e}^2)\epsilon_\ell \,, \qquad
s_{R13}\simeq \left|\frac{C^{y\,*}_\Delta C^y_V+C^{y\,*}_{V\Delta}C^y}{C^{y\, 2}}\right |
s_e \delta'_e \epsilon_\ell  \,,
\end{align}
respectively.
The phase  $\phi_{L13}$ and $\phi_{R13}$ are:
\begin{align}
\phi_{L13}\simeq \arg \frac{C^y_{\Delta}C^{y\,*}_{V\Delta}}{|C^{y}|^2}  \,, \qquad\qquad
\phi_{R13}\simeq 
0 \,.
\label{U13}
\end{align}
It is noticed that  the  $(1,3)$ and $(2,3)$ components
of $(Y^\dagger Y)_{RR}$ have the same phase as seen in Eq.\eqref{Y2RR}.
Therefore,   the phase matrix $P_{R23}$  removes phases of both $(1,3)$ and $(2,3)$ components, that is,  $\phi_{R13}\simeq 0$ is derived.


In mass-eigenstate basis, the  matrix $X^{e\gamma}$ in Eq.\,\eqref{U2egamma}
is transformed by the unitary matrix of Eq.\,\eqref{unitary} 
as $U_L^\dagger X^{e\gamma} U_R$, 
whose elements
correspond to the Wilson coefficients
of Eq.\,\eqref{dipole-operators} :
{
\begin{align}
&\CL{e\gamma}{ee}^\prime = [ U_L^{ \dag} X^{e \gamma}  U_R ]_{11}
\simeq |C^{e\gamma}_\Delta|  \delta_e' \left[ \frac{1}{c_e}
+\frac{s_e^2}{c_e}\epsilon_{\ell}^2 
 \left( \frac{C^{e \gamma}_{VV\Delta}}{C^{e \gamma}_\Delta}
  - \frac{C^{y}_{VV\Delta}}{C^{y}_\Delta} 
 \right) \right] +\delta_e'\frac{s_e^2}{c_e}\epsilon_{\ell}^2 \, C_{ 3rd}\,,
 \nonumber \\
&\CL{e\gamma}{\mu\mu}^\prime =[U_L^{ \dag} X^{e \gamma} 
 U_R ]_{22} \nonumber\\
&\quad\ \simeq |C^{e\gamma}_\Delta|  \delta_e \left[ \frac{1}{c_e}
	+
c_e \epsilon_{\ell}^2
\left( \left |\frac{C^{e \gamma}_{VV\Delta}}{C^{e \gamma}_\Delta}\right|
	\cos{(\arg{C^{e \gamma}_{VV\Delta}})} -  \left|\frac{C^{y}_{VV\Delta}}{C^{y}_\Delta}\right|
	\frac{s_e^2}{c_e^2}\cos{(\arg{C^y_{VV\Delta}})}
 \right)  \right] \nonumber\\
&\quad\ + i
|C^{e\gamma}_\Delta| c_e \delta_e\epsilon_{\ell}^2
\left (\left |\frac{C^{e \gamma}_{VV\Delta}}{C^{e \gamma}_\Delta}\right|
\sin{(\arg{C^{e \gamma}_{VV\Delta}})} -  \left|\frac{C^{y}_{VV\Delta}}{C^{y}_\Delta}\right|
\sin{(\arg{C^y_{VV\Delta}})}\right )
	+ c_e \delta_e\epsilon_{\ell}^2\, C_{3rd}\,,
\nonumber\\
&\CL{e\gamma}{e\mu}^\prime =[ U_L^{ \dag} X^{e \gamma}  U_R ]_{12} 
\simeq |C^{e\gamma}_\Delta|  s_e \delta_e \epsilon_{\ell}^2 
	\left( \frac{C^{e \gamma}_{VV\Delta}}{C^{e \gamma}_\Delta}
 - \frac{C^{y}_{VV\Delta}}{C^{y}_\Delta}
\right)
+
s_e \delta_e \epsilon_{\ell}^2 \, C_{3rd}\,,\nonumber \\
&\CL{e\gamma}{\mu e}^\prime =[ U_L^{ \dag} X^{e \gamma}  U_R ]_{21}
\simeq |C^{e\gamma}_\Delta|  s_e \delta'_e \epsilon_{\ell}^2 
\left( \frac{C^{e \gamma}_{VV\Delta}}{C^{e \gamma}_\Delta}
 -  \frac{C^{y}_{VV\Delta}}{C^{y}_\Delta}\right)
+s_e \delta'_e \epsilon_{\ell}^2 \, C_{3rd}\,, 
\label{Wilson12}
\end{align}
}
where $C_{3rd}$ is given in terms of third-family parameters as:
\begin{align}
C_{3rd}=\frac{C^{y*}_\Delta C^y_V+C^{y*}_{V\Delta}C^y}{|C^y|^2}
	 \left ( \frac{C^{e\gamma}}{C^y}C^y_V -C^{e\gamma}_V\right )+
\frac{C^y_V}{C^y} 
\left (\frac{C^{e\gamma}_V}{C^y_V}C^y_{V\Delta} -C^{e\gamma}_{V\Delta}  \right )	 \,.
\end{align}
The contribution $C_{3rd}$ was not taken account in the analysis of Ref.\cite{Isidori:2021gqe}.
It is found that the $C_{3rd}$ 
 is of order 1 and  does not vanish unless
$C^{e\gamma}_V=C^y_V$, $C^{e\gamma}_{V\Delta}=C^y_{V\Delta}$
and $C^{e\gamma}=C^y$.
It is remarked that   the off-diagonal component of
 $\CL{e\gamma}{e\mu(\mu e)}^\prime$ are not suppressed due to  $C_{3rd}$ even if
the alignment 
$C^{e\gamma}_{V\Delta}/C^{e\gamma}_\Delta
=C^{y}_{V\Delta}/C^{y}_\Delta$ is imposed \cite{Isidori:2021gqe}.
The third family contribution on the $\mu \to e\gamma$ decay
is comparable to the one of the first- and second-family.
It is also remarked that 
 the non-vanishing electron EDM is
  realized from the third family even if there is no CP phase in first- and second-family  as seen in the imaginary part of
   $\CL{e\gamma}{e e}^\prime $ in  Eq.\,\eqref{Wilson12}.

  The Wilson coefficients with respect  to the third-family are given as:
 \begin{align}
  &\CL{e\gamma}{e\tau}^\prime = [ U_L^{ \dag} X^{e \gamma}  U_R ]_{13}
 \simeq 
 \left ( {C^{e\gamma}_{V}- \frac{C^{e\gamma}}{C^{y}} C^{y}_{V}} \right )
 \frac{s_e}{c_e}\epsilon_{\ell} \,,
 \nonumber \\
  &\CL{e\gamma}{\tau e}^\prime = [ U_L^{ \dag} X^{e \gamma}  U_R ]_{31}
 \simeq 
 \left ( {C^{e\gamma}_{V\Delta}- \frac{C^{e\gamma}}{C^{y}} C^{y}_{V\Delta}} \right )
 s_e \delta'\epsilon_{\ell} \,,
 \nonumber \\
 &\CL{e\gamma}{\mu\tau}^\prime = [ U_L^{ \dag} X^{e \gamma}  U_R ]_{23}
\simeq \left ( {C^{e\gamma}_{V}- \frac{C^{e\gamma}}{C^{y}} C^{y}_{V}} \right ) \epsilon_{\ell}   \,,
\nonumber \\
  &\CL{e\gamma}{\tau\mu}^\prime = [ U_L^{ \dag} X^{e \gamma}  U_R ]_{32}
 \simeq  \left ( {C^{e\gamma}_{V\Delta}- \frac{C^{e\gamma}}{C^{y}} C^{y}_{V\Delta}} \right ) c_e \delta\epsilon_{\ell}  \,,
 \nonumber \\
 &\CL{e\gamma}{\tau\tau}^\prime = [ U_L^{ \dag} X^{e \gamma}  U_R ]_{33}
 \simeq C^{e\gamma} +
 C^{e \gamma}_V \frac{C^y_V}{C^{y}}  
 \epsilon_{\ell}^2   \,.
 \label{Wilson23-13}
 \end{align}

As seen in Eq.\,\eqref{Wilson12}, 
${\rm Im}\, \CL{e\gamma}{ee}^\prime$, which gives  the electron EDM,
is suppressed by the factor of $s_e^2\epsilon_{\ell}^2\delta_e'/\delta_e$
while $\mu\to e\gamma$  ($| \CL{e\gamma}{e\mu}^\prime|$) is suppressed by $s_e\epsilon_{\ell}^2$
compared with the muon $(g-2)_\mu$.
It is noted that the $\tau_R \to e_L \gamma$ transition is
not so suppressed
due to the  $e_L$-$\mu_L$ mixing angle   $s_e=0.01-0.1$ as seen in Eq.\,\eqref{Wilson23-13} although $(X^{e \gamma})_{13}$ vanishes in the flavor basis as seen in Eq.\,\eqref{U2egamma}

In the next section, we show the numerical results without approximations.
 Indeed, it is found that  the approximate Wilson coefficients in Eqs.\,\eqref{Wilson12} and  \eqref{Wilson23-13} are agree with  exact ones in accuracy of  $10\,\%$  numerically.

\section{Numerical analyses}
\label{numerical}

Let us discuss Wilson coefficients numerically in mass-eigenvalue basis. 
 The parameters $\delta_e$ and $\delta_e'$ are determined
  according to the observed charged lepton mass ratios, $m_e/m_\mu$ and  $m_\mu/m_\tau$
   as in Eq.\,\eqref{massratio}.
  On the other hand, 
 we scan $s_e$ and $\epsilon_\ell$ in the region of Eq.\,\eqref{s-ep},
   $[0.01-0.1]$  with equal weights,
   that is the random scan in the linear space.
 As seen in Eqs.\eqref{Wilson12} and \eqref{Wilson23-13}, the leading terms of off-diagonal Wilson coefficients  depend on these two parameters, while the leading terms of the diagonal ones do not. Therefore, the predictions of the LFV processes depend crucially on the scan range of these parameters. Since the imaginary parts of the diagonal Wilson coefficients also depend on these two parameters, the predictions of the EDM depend on the scan range of these parameters.
 The other parameters are 
  $C^y$, $C^y_{\Delta}$, $C^y_{V}$, $C^y_{V\Delta}$
  $C^y_{VV\Delta}$, 
   $C^{e\gamma}$, $C^{e\gamma}_{\Delta}$, $C^{e\gamma}_{V}$, $C^{e\gamma}_{V\Delta}$ and 
  $C^{e\gamma}_{VV\Delta}$ 
  which  are  complex parameters of order $1$.
   Since the relative phases contribute to the observables except for EDM,
    we take $C^y_{\Delta}$ and $C^{e\gamma}_{\Delta}$ to be real
     in order to predict the electron EDM properly removing unphysical CP phases.
 Those parameters are put in
  the normal distribution with an average $1$ and standard deviation 
  $0.25$, which is given
   in the work of the flavor structure of quarks and leptons
   with reinforcement learning \cite{Nishimura:2023wdu}.
   (We have checked that our numerical results are not so changed 
  even if the standard deviation $0.5$ is taken.)
   The phases are scanned  to be random in $0\sim 2\pi$.
   
   Then,  we construct the matrices 
   Eqs.\,\eqref{U2y} and \eqref{U2egamma} numerically.
   After diagonalizing Yukawa matrices in Eq.\eqref{U2y},
   we can obtain  $\CL{e\gamma}{\alpha\beta}^\prime$'s
    in the mass-eigenvalues basis.
    In the following analysis, we have only taken the regions where $\text{Re}\  [\CL{e\gamma}{\mu\mu}^\prime] =1.0 \times 10^{-5}$, corresponding to the $(g-2)_{\mu}$ anomaly, can be reproduced.
It is noted that the $U(2)$ flavor symmetry is supposed 
in the electroweak scale.  Therefore,
 the contribution on the leptonic dipole operator of the renormalization group equations (RGEs)
 is neglected (see also the study in the  subsection 4.3 of  Ref.\cite{Kobayashi:2021pav}.).

\subsection{Prediction of $(g-2)_{e}$ and $(g-2)_{\tau}$}

The NP effect in $(g-2)_{\ell}$ is occurred by the diagonal
components of the Wilson coefficient of the leptonic dipole operator
at  mass-eigenstate basis.
We have the ratios of the diagonal coefficients from
Eqs.\,\eqref{Wilson12} and \eqref{Wilson23-13} as:
\begin{eqnarray}
\label{CG-ee}
&&\frac{  |\CL{e\gamma}{e e}^\prime  |}
{|\CL{e\gamma}{\mu \mu}^\prime  |} \simeq \frac{\delta_e'}{\delta_{e}} \simeq \frac{m_e}{m_\mu}
\simeq 4.8\times 10^{-3} \,,
\qquad\quad 
\frac{  |\CL{e\gamma}{\mu\mu}^\prime  |}
 { | \CL{e\gamma}{\tau\tau}^\prime |} =
  \frac{ |C^{e\gamma}_\Delta |}{| C^{e\gamma}|} \frac{1}{c_e}\delta_e
\simeq \frac{m_\mu}{m_\tau}\simeq  5.9\times 10^{-2} 
\,,
\label{CGratio}
\end{eqnarray}
where the second  equation is derived by taking account of 
$|C^{e\gamma}_\Delta |\sim| C^{e\gamma}|\sim 1$ and $c_e\sim 1$.
Inputting the observed charged lepton masses with $3\,\sigma$ error-bar, we have obtained 
\begin{align}
\delta_e=(5.0-6.0)\times  10^{-2} \,,\qquad\quad \delta'_e=(2.3-3.0)\times 10^{-4}\,,
\label{delta}
\end{align}
in the numerical results.

Suppose that  the leptonic  dipole operator is responsible for  the observed anomaly of
$(g-2)_{\mu}$.
Inputting the reference value of Eq.\,\eqref{muon-input}, we can estimate the magnitude of  the electron $(g-2)_e$ 
by using the  relation in Eq.\,\eqref{CGratio} as:
\begin{align}
\Delta a_{e} &= \frac{4 m_e}{e}   \frac{v}{\sqrt 2}\,\frac{1}{\Lambda^2} 
\text{Re} \,[\CL{e\gamma}{ee}^\prime] \simeq  6.2\times 10^{-14}\,,
\end{align}
where $v\approx 246$~GeV.
It is easily seen that $\Delta a_{e}$ 
and  $\Delta a_{\mu}$ are proportional to the lepton {masses squared}.
This result is agreement with the naive scaling $\Delta a_\ell \propto m^2_\ell$\cite{Giudice:2012ms}.

In the electron {anomalous} magnetic moment, 
the  experiments \cite{Hanneke:2008tm} give
\begin{align}
a_e^\mathrm{Exp}=1\, 159\, 652\, 180.73(28)\times 10^{-12}\,,
\end{align}
while 
the SM prediction crucially depends on the input value for the fine-structure constant $\alpha$.  
The two latest determinations
\cite{Parker:2018vye,Rb:2020} based on 
Cesium and Rubidium atomic recoils differ by more than $5\sigma$.
Those observations lead to the difference {from the} SM prediction
\begin{align}
&&\Delta a_e^{Cs} &= a_e^\mathrm{Exp} - a_e^\mathrm{SM,CS} = \brackets{-8.8 \pm 3.6} \times 10^{-13}~\,, \nonumber\\
&&\Delta a_e^{Rb} &= a_e^\mathrm{Exp} - a_e^\mathrm{SM,Rb} = \brackets{4.8 \pm 3.0} \times 10^{-13}~\,.
\end{align}
Our predicted value is small of one order
compared with the observed one at present. 
We wait for the precise determination of the fine structure constant
to test our flavor model.

The  tauon $(g-2)_\tau$  is also predicted 
by using the  relation in Eq.\eqref{CGratio} as:
\begin{align}
\Delta a_{\tau} &= \frac{4 m_\tau}{e}   \frac{v}{\sqrt 2}\,\frac{1}{\Lambda^2} 
\text{Re} \,[\CL{e\gamma}{\tau\tau}^\prime] \simeq  7.5\times 10^{-7}\,,
\end{align}
which is also proportional to the lepton masses squared.

\subsection{$(g-2)_{\mu}$ versus $\mu\to e\gamma$, $\tau\to\mu\gamma$
and $\tau\to e\gamma$}
The NP in the LFV  process
is severely constrained by 
the experimental  bound of the $\mu\to e\gamma$ decay
 in the MEG experiment~\cite{TheMEG:2016wtm}.
It is also constrained by the $\tau\to \mu\gamma$ decay
  in the Belle experiment \cite{Belle:2021ysv}. 
We can discuss  the correlation 
between the anomaly of  the muon $(g-2)_\mu$ and
the LFV process $\mu\to e\gamma$ by using the Wilson coefficients
in Eq.\,\eqref{Wilson12}.
The  ratios are given as:
\begin{eqnarray}
\left |\frac{ \CL{e\gamma}{e \mu}^\prime}{\CL{e\gamma}{\mu \mu}^\prime}
\right |
\simeq  c_e s_e \epsilon_\ell^2
	\left | \frac{C^{e \gamma}_{VV\Delta}}{C^{e \gamma}_\Delta}
 -  \frac{C^{y}_{VV\Delta}}{C^{y}_\Delta}
 + \frac{C_{3rd}}{C^{e \gamma}_\Delta} \right |
 \,, 
\label{muReL}
\end{eqnarray}
and
\begin{eqnarray}
\left |\frac{ \CL{e\gamma}{\mu e}^\prime}{\CL{e\gamma}{\mu \mu}^\prime}
\right |
\simeq  c_e s_e \frac{\delta'_e}{\delta_e}\epsilon_\ell^2
\left | \frac{C^{e \gamma}_{VV\Delta}}{C^{e \gamma}_\Delta}
-  \frac{C^{y}_{VV\Delta}}{C^{y}_\Delta} 
+ \frac{C_{3rd}}{C^{e \gamma}_\Delta} \right |
 \,. 
\label{muLeR}
\end{eqnarray}

We also obtain ratios for $\tau\to\mu\gamma$ and $\tau\to e\gamma$ processes as:
\begin{eqnarray}
\left |\frac{ \CL{e\gamma}{\mu\tau}^\prime}{\CL{e\gamma}{\mu \mu}^\prime}
\right |
\simeq  c_e  \epsilon_\ell
 \frac{1}{\delta_e}\frac{1}{|C^{e\gamma}_\Delta|}
\left | C^{e \gamma}_V
-  \frac{C^{e\gamma}}{C^{y}} C^y_V\right | \,,
\qquad
\left |\frac{ \CL{e\gamma}{\tau\mu}^\prime}{\CL{e\gamma}{\mu \mu}^\prime}
\right |
\simeq  c_e^2  \epsilon_\ell \frac{1}{|C^{e\gamma}_\Delta|}
\left | C^{e \gamma}_V
-  \frac{C^{e\gamma}}{C^{y}} C^y_V\right |\,,
\label{taumu}
\end{eqnarray}
and 
\begin{eqnarray}
\left |\frac{ \CL{e\gamma}{e\tau}^\prime}{\CL{e\gamma}{\mu\mu}^\prime}
\right |
\simeq  s_e  \epsilon_\ell 
\frac{1}{\delta_e}\frac{1}{|C^{e\gamma}_\Delta|}
\left | C^{e \gamma}_V
-  \frac{C^{e\gamma}}{C^{y}} C^y_V\right | \,,
\qquad
\left |\frac{ \CL{e\gamma}{\tau  e}^\prime}{\CL{e\gamma}{\mu \mu}^\prime}
\right |
\simeq  c_e s_e  \frac{\delta'_e}{\delta_e}\epsilon_\ell \frac{1}{|C^{e\gamma}_\Delta|}
\left | C^{e \gamma}_V
-  \frac{C^{e\gamma}}{C^{y}} C^y_V\right | \,.
\label{taue}
\end{eqnarray}

 For the $\tau_R\to \mu_L\gamma$ process,  the ratio in Eq.\,\eqref{taumu} is proportional to 
 $\epsilon_\ell/\delta_e$, which is expected to be of order 1.
 Since $\delta_e$ is  fixed in  Eq.\,\eqref{delta},
  we have a strong constraint for $\epsilon_{\ell}$
   from the experimental upper-bound
 of the Wilson coefficient  in Eq.\,\eqref{eq:bound_C_egamma_23-13}.
 On the other hand, 
 for  the $\mu_R\to e_L\gamma$ process, the ratio in Eq.\,\eqref{muReL}  is proportional to  $s_e\epsilon_\ell^2$.
 Therefore, $s_e$ is also constrained considerably as well as $\epsilon_{\ell}$.

\begin{figure}[t]
	\begin{minipage}[]{0.48\linewidth}
		\includegraphics[{width=\linewidth}]{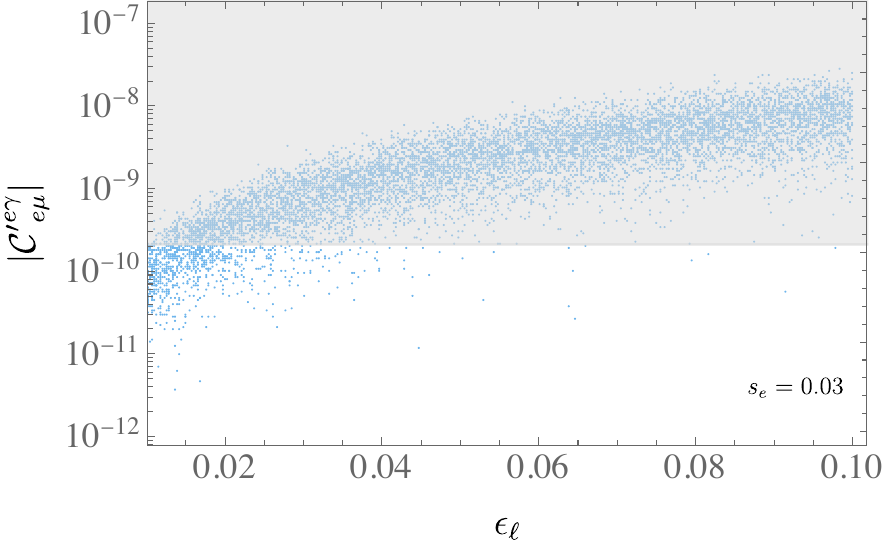}
		\caption{$|\CL{e\gamma}{e \mu}^\prime|$
		 in $1/\Lambda^2\  [\mathrm{TeV}^{-2}]$ unit versus $\epsilon_{\ell}$,
			where $s_e=0.03$ is put and the grey region is excluded by the $\mu\to e\gamma$ experiment.}
		\label{fig:muReLgamma}
	\end{minipage}
	\hspace{5mm}
	\begin{minipage}[]{0.48\linewidth}
		\includegraphics[{width=\linewidth}]{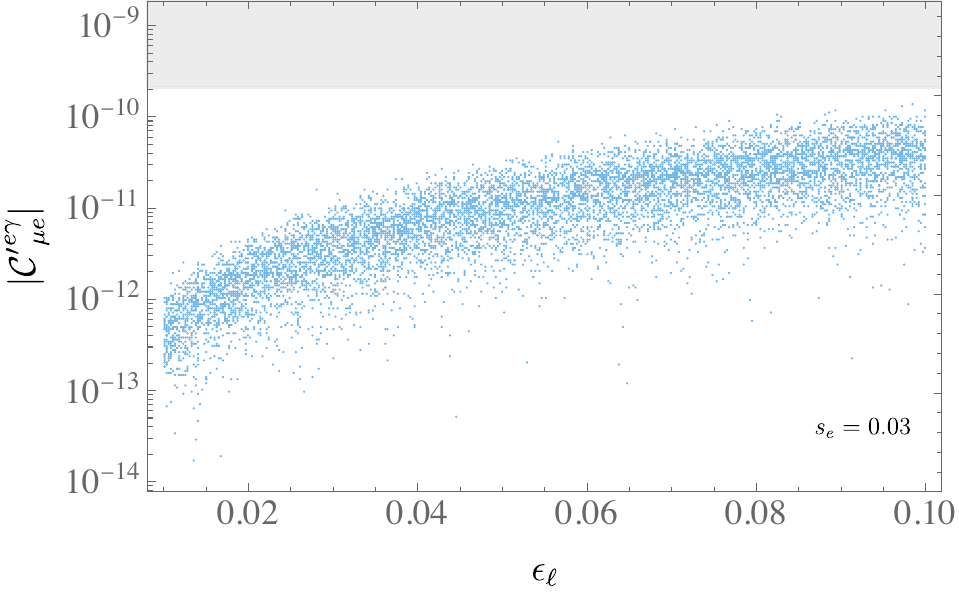}
		\caption{ $|\CL{e\gamma}{\mu e}^\prime|$
	 in  $1/\Lambda^2\  [\mathrm{TeV}^{-2}]$ unit versus $\epsilon_{\ell}$,
	where $s_e=0.03$ is put and the grey region is excluded by the $\mu\to e\gamma$ experiment.}
\label{fig:muLeRgamma}
	\end{minipage}
\end{figure}

 In order to see the predicted  range of $\mu\to e \gamma$,
 we plot the magnitude of $\CL{e\gamma}{e\mu}^\prime$
 versus $\epsilon_{\ell}$, which corresponds to $\mu_R\to e_L\gamma$ process,
 in Fig.\,\ref{fig:muReLgamma}.
  In this figure, $s_e=0.03$ is put as a benchmark.
  It is found that $\epsilon_{\ell}$ should be smaller than around  $0.05$
  in order to obtain $\CL{e\gamma}{e\mu}^\prime$ below the experimental
   upper-bound. The expected value of the branching ratio 	$\mathcal{B}({\mu \to e \gamma})$
   will be discussed in  subsection \ref{frequency}.

 We also show the magnitude of $\CL{e\gamma}{\mu e}^\prime$
 versus $\epsilon_{\ell}$ with $s_e=0.03$,
 which  corresponds $\mu_L\to e_R\gamma$ process
 in Fig.\,\ref{fig:muLeRgamma}. 
 The magnitude of $\CL{e\gamma}{\mu e}^\prime$ is  suppressed 
 compared with $\CL{e\gamma}{e\mu}^\prime$.
 It is remarked  that the NP signal of the $\mu\to e\gamma$ process 
mainly comes from the  operator $\bar e_L \sigma_{\mu\nu}\mu_R$
in the $U(2)$ flavor model.  
Indeed, the ratio is given as:
\begin{eqnarray}
\left |\frac{ \CL{e\gamma}{\mu e}^\prime}{\CL{e\gamma}{e \mu}^\prime}
\right |
\simeq \frac{\delta_e'}{\delta_e}\simeq \frac{m_e}{m_\mu} \,.
\label{massratio12}
\end{eqnarray}
The angular distribution with respect to the muon polarization can distinguish between $\mu_R \to e_L\gamma$ and $\mu_L
\to e_R\gamma$
\cite{Okada:1999zk}.

\begin{figure}[t]
	\begin{minipage}[]{0.48\linewidth}
		\includegraphics[{width=\linewidth}]{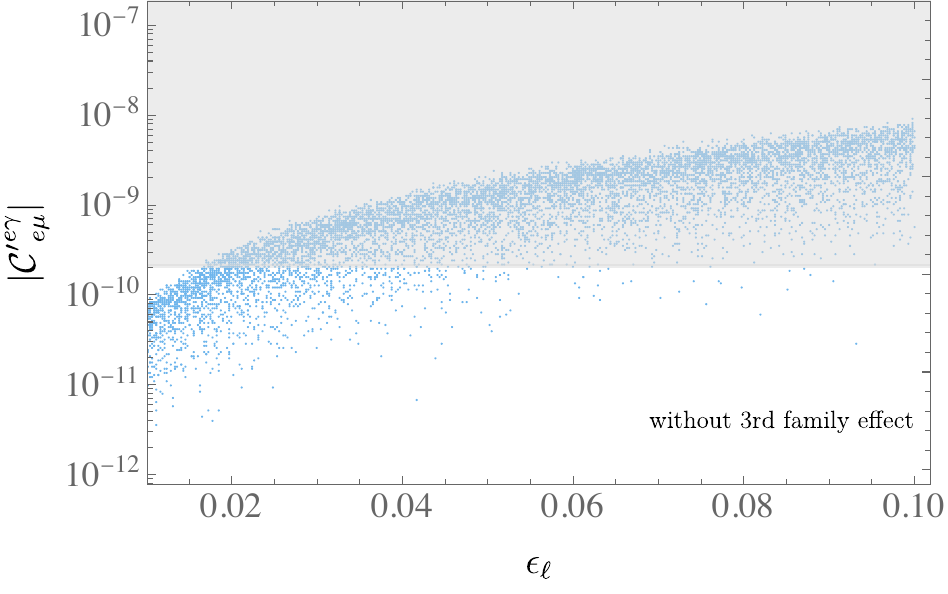}
		\caption{$|\CL{e\gamma}{e \mu}^\prime|$
			 in  $1/\Lambda^2\  [\mathrm{TeV}^{-2}]$ unit versus  $\epsilon_{\ell}$
			for case (a) (without third-family contribution),
			where $s_e=0.03$ is put and the grey region is excluded by the experiment of $\mu\to e\gamma$.}
		\label{fig:muReLgamma12}
	\end{minipage}
	\hspace{5mm}
	\begin{minipage}[]{0.48\linewidth}
		\includegraphics[{width=\linewidth}]{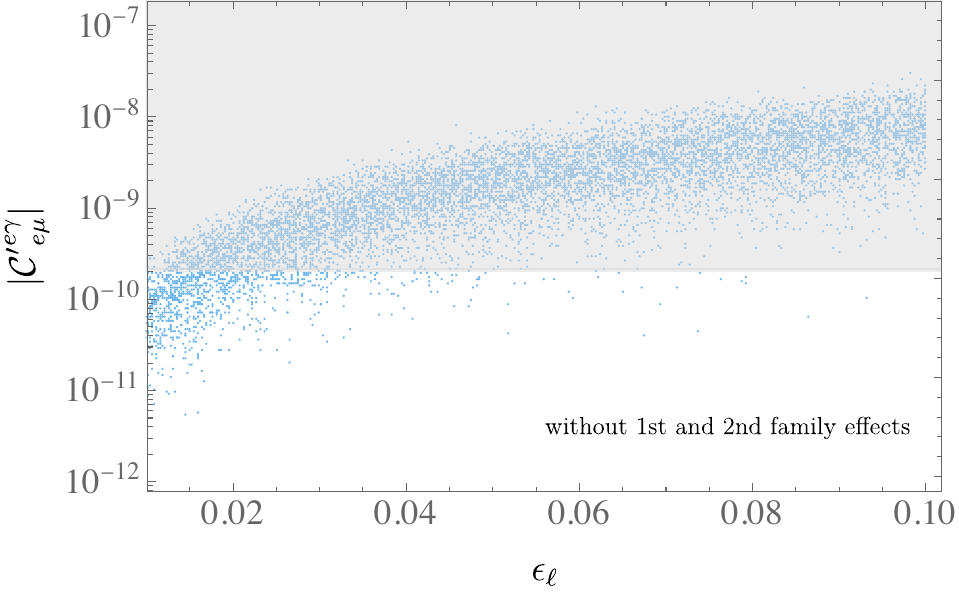}
		\caption{$|\CL{e\gamma}{e \mu}^\prime|$
			in $1/\Lambda^2\  [\mathrm{TeV}^{-2}]$ unit
			 versus  $\epsilon_{\ell}$
		for case (b) (without first- and second-family contribution),
			where $s_e=0.03$ is put and the grey region is excluded by the experiment of $\mu\to e\gamma$.}
		\label{fig:muReLgamma3}
	\end{minipage}
\end{figure}

In order to see the contribution of the third-family 
to  $\CL{e\gamma}{e \mu}^\prime$, we show 
the magnitude of $\CL{e\gamma}{e \mu}^\prime$
for both cases of alignment of coefficients
(a) $C^{e\gamma}_V=C^y_V$, $C^{e\gamma}_{V\Delta}=C^y_{V\Delta}$, $C^{e\gamma}=C^y$ and
(b) $C^{e\gamma}_{V\Delta}/C^{e\gamma}_\Delta
=C^{y}_{V\Delta}/C^{y}_\Delta$.
The case  (a) corresponds to the case  excluding the third-family contribution
and (b) to the case excluding the first- and second-family contribution.
The numerical results are presented
in Fig.\,\ref{fig:muReLgamma12} for case (a) and Fig.\,\ref{fig:muReLgamma3} 
for case (b).
Thus, the third-family contribution is comparable or
 rather large compared with the one from  the first- and 
 second-family.

\begin{figure}[t!]
	\begin{minipage}[]{0.48\linewidth}
		\includegraphics[{width=\linewidth}]{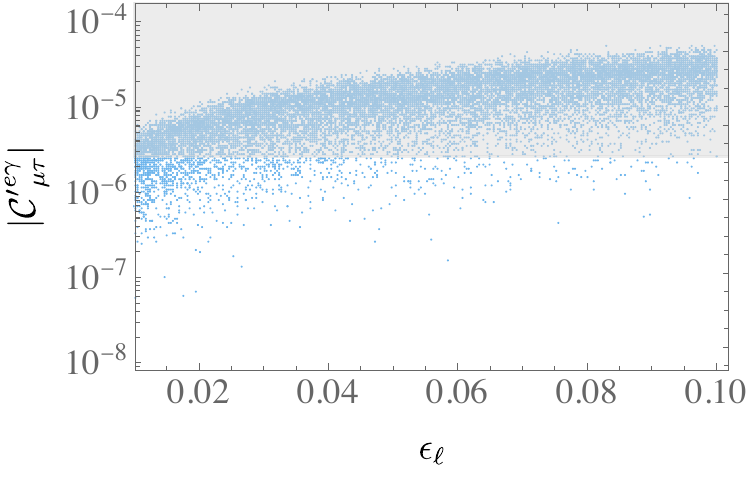}
		\caption{$|\CL{e\gamma}{\mu \tau}^\prime|$
		in $1/\Lambda^2\  [\mathrm{TeV}^{-2}]$ unit versus $\epsilon_{\ell}$,
			where  the grey region is excluded by the $\tau\to \mu\gamma$ experiment.}
		\label{fig:ep-muLtauR}
	\end{minipage}
	\hspace{5mm}
	\begin{minipage}[]{0.48\linewidth}
		\includegraphics[{width=\linewidth}]{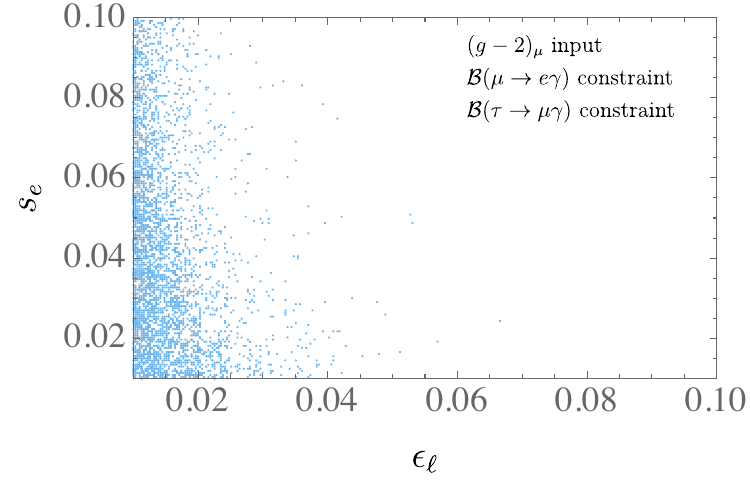}
		\caption{Allowed region in the  $\epsilon_{\ell}$-$s_e$ plane
		by the upper-bound of
		 $\mu\to e\gamma$ and  $\tau\to \mu\gamma$ branching ratios. }
		\label{fig:epsilon-s}
	\end{minipage}
\end{figure}

The $\tau\to\mu\gamma$ decay is an interesting process
in the $U(2)$ flavor symmetry 
because it is  suppressed by only $\epsilon_{\ell}$. \cite{Isidori:2021gqe}.
The upper-bound of the branching ratio of $\tau\to\mu\gamma$
constrains the magnitude of  $\epsilon_{\ell}$ 
as discussed below  Eq.\,\eqref{taue}.
We show the magnitude of $\CL{e\gamma}{\mu\tau}^\prime$ 
versus $\epsilon_{\ell}$ in Fig.\,\ref{fig:ep-muLtauR},
where the predicted value is almost independent of $s_e$
as seen in Eq.\,\eqref{taumu}.
It is found that small $\epsilon_{\ell}$ (much less than $0.1$) is  favored
 by the experimental  upper-bound of the $\tau\to\mu\gamma$ decay.
On the other hand, $\CL{e\gamma}{\tau\mu}^\prime$ is suppressed
in  more than one order by $\delta_e$. We omit a figure for this process.

\begin{figure}[t]
	\begin{minipage}[]{0.47\linewidth}
		\includegraphics[{width=\linewidth}]{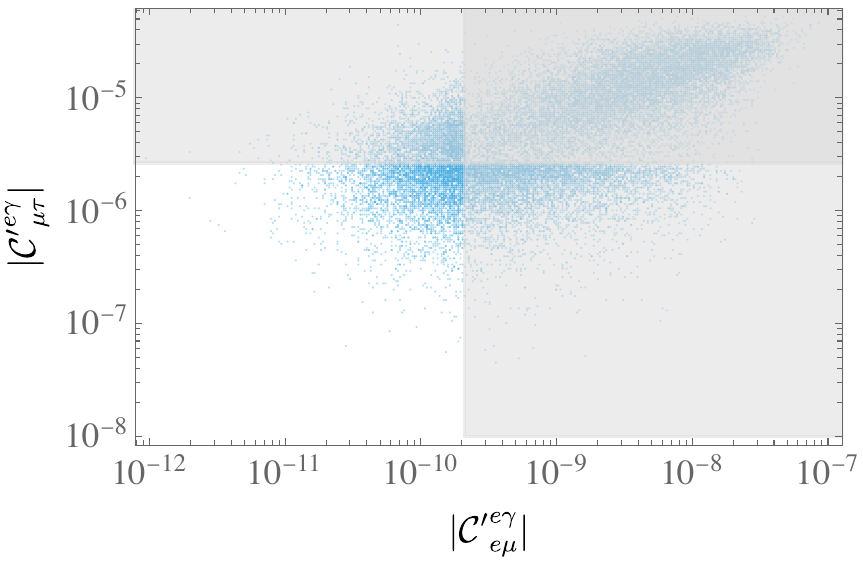}
		\caption{ $|\CL{e\gamma}{\mu \tau}^\prime|$
			in $1/\Lambda^2\  [\mathrm{TeV}^{-2}]$ unit versus
			$|\CL{e\gamma}{e \mu}^\prime|$.
			The grey region is excluded by the experiments.}
		\label{fig:eLmuR-muLtauR}
	\end{minipage}
	\hspace{5mm}
	\begin{minipage}[]{0.47\linewidth}
		\includegraphics[{width=\linewidth}]{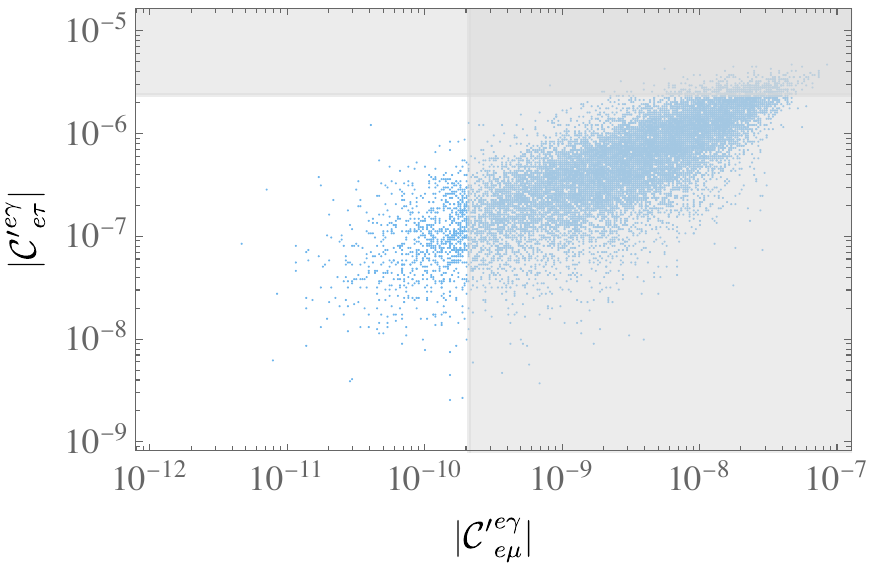}
		\caption{$|\CL{e\gamma}{e \tau}^\prime|$ 
			in $1/\Lambda^2\  [\mathrm{TeV}^{-2}]$ unit versus
			$|\CL{e\gamma}{e \mu}^\prime|$.
		The grey region is excluded by the experiments.}
		\label{fig:muLtauR-eLtauR}
	\end{minipage}
\end{figure}

As seen in Figs.\,\ref{fig:muReLgamma} and \ref{fig:ep-muLtauR},
the substantial regions of predicted Wilson coefficients exceed the experimental upper-bounds.
Imposing the  upper-bounds of the branching ratios of
$\mu\to e\gamma$ and  $\tau\to \mu\gamma$ decays,
we obtain the allowed region 
in $\epsilon_\ell-s_e$ plane  in Fig.\,\ref{fig:epsilon-s}.
It is found that $\epsilon_{\ell}$ is almost smaller than $0.05$
while $s_e$ is allowed in rather wide range
of $[0.01-0.1]$. 


In order to see the correlation between 
$|\CL{e\gamma}{e \mu}^\prime|$ and $|\CL{e\gamma}{\mu \tau}^\prime|$,
we plot their predictions  in Fig.\,\ref{fig:eLmuR-muLtauR}.
It is found  that the allowed region is restricted.
In the next subsection, we discuss the expectation value of
$\mu\to e\gamma$ and $\tau\to \mu\gamma$. 

We also show the predicted region of $|\CL{e\gamma}{e \tau}^\prime|$ 
versus $|\CL{e\gamma}{\mu \tau}^\prime|$ in Fig.\,\ref{fig:muLtauR-eLtauR}.
The magnitude of  $\CL{e\gamma}{e \tau}^\prime$ is  proportional to $|\CL{e\gamma}{e \mu}^\prime|$ roughly.
The prediction of $\CL{e\gamma}{e \tau}^\prime$ is still low in one order compared with the  experimental upper-bound.
It is noted that $\CL{e\gamma}{\tau e }^\prime$ is suppressed in 
$\delta'_e$ as seen in Eq.\,\eqref{Wilson23-13}.
We omit a figure for this process.

\subsection{ Predictions of EDM }

In the allowed region of $\epsilon_\ell$ and $s_e$ in Fig.\,\ref{fig:epsilon-s},
we  discuss  the EDM of the charged leptons.
The electron EDM comes from the imaginary part of 
$\CL{e\gamma}{e e}^\prime$, and the magnitude is estimated approximately from Eq.\,\eqref{Wilson12}.
 The ratio of ${\rm Im}\ \CL{e\gamma}{e e}^\prime$ and
  ${\rm Re}\ \CL{e\gamma}{\mu \mu}^\prime$ is bounded by the observed constraints in   
    Eqs.\,\eqref{Cmumu-exp} and \eqref{Cee-exp} as:
\begin{eqnarray}
\left |\frac{ {\rm Im}\ \CL{e\gamma}{e e}^\prime}
{{\rm Re}\ \CL{e\gamma}{\mu \mu}^\prime}
\right |
\simeq s_e^2 \frac{\delta'}{\delta} \epsilon_\ell^2\,
{\rm Im} \, \left ( \frac{C^{e \gamma}_{VV\Delta}}{C^{e \gamma}_\Delta}
-  \frac{C^{y}_{VV\Delta}}{C^{y}_\Delta}+
\frac{C_{3rd}}{|C^{e \gamma}_\Delta|} \right )
 < 1.8\times 10^{-8} \,.
\end{eqnarray}
 Putting $\epsilon_{\ell}=0.03$ and $s_e=0.03$
 with $\delta'_e/\delta_e\simeq m_e/m_\mu$ in Eq.\,\eqref{massratio12},
the factor in front of the middle equation,  $s_e^2 \epsilon_\ell^2 \delta'/\delta $, is  
 $4\times 10^{-9}$. Thus, the ratio is possibly predicted
 to be ${\cal O}(10^{-8})$, which corresponds to 
 $|d_e|\sim 10^{-30}\, {\rm e\,cm}$.
  We expect it to be  detectable in the near future.
  
\begin{figure}[t]
	\begin{minipage}[]{0.47\linewidth}
		\includegraphics[{width=\linewidth}]{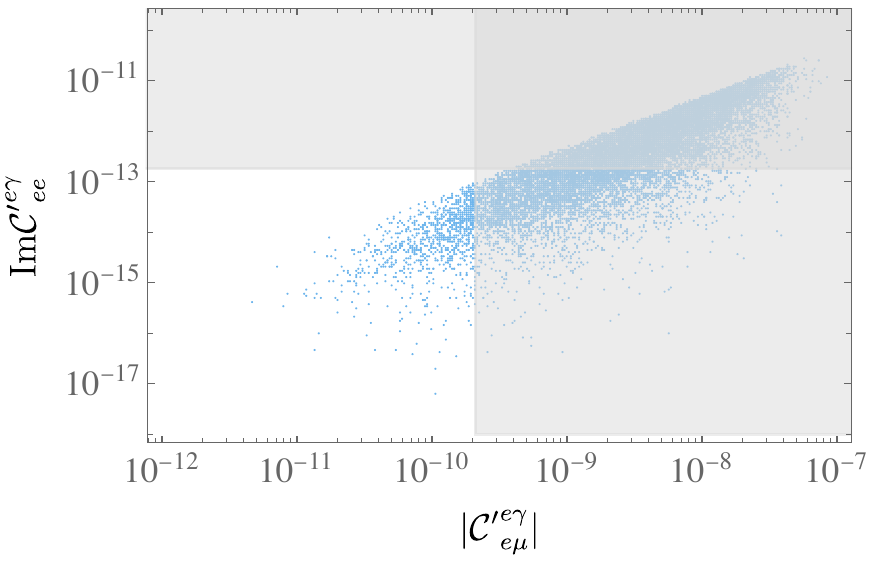}
		\caption{${\rm Im}\ \CL{e\gamma}{e e}^\prime$ versus
		$ |\CL{e\gamma}{e \mu}^\prime|$
		(electron EDM versus $\mu\to e\gamma$)
			in $1/\Lambda^2\  [\mathrm{TeV}^{-2}]$ unit.
			The grey region is excluded by the experiments.}
		\label{fig:eEDM123}
	\end{minipage}
	\hspace{5mm}
	\begin{minipage}[]{0.47\linewidth}
		\includegraphics[{width=\linewidth}]{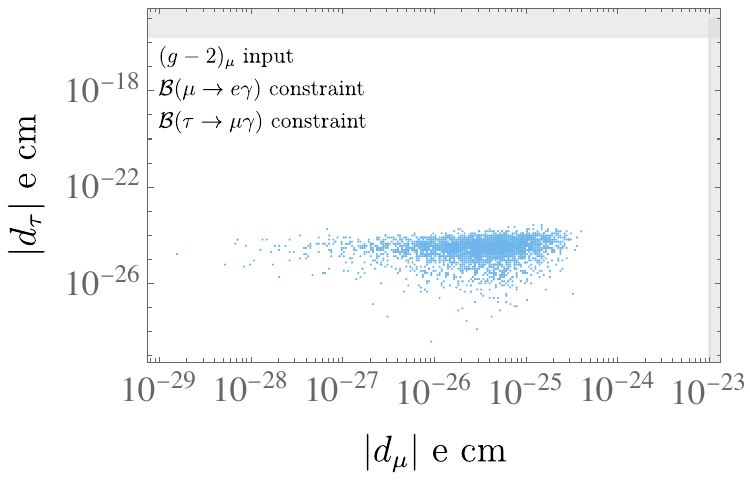}
		\caption{Predicted  EDMs of  muon and tauon in $[\rm e\,cm]$ unit.
				The grey region is excluded by the experiments.
	}
		\label{fig:mutauEDM}
	\end{minipage}
\label{mue}
\end{figure}

Fig.\,\ref{fig:eEDM123} shows a plot of 
${\rm Im}\ \CL{e\gamma}{e e}^\prime$ versus $\CL{e\gamma}{e \mu}^\prime$.
These ratio can be expressed approximately as :
\begin{eqnarray}
\frac{ {\rm Im}\ \CL{e\gamma}{e e}^\prime}
{|\CL{e\gamma}{e \mu}^\prime|}
\simeq \frac{s_e}{c_e}\frac{\delta'}{\delta}
\frac{
	{\rm Im} \, \left ( \frac{C^{e \gamma}_{VV\Delta}}{C^{e \gamma}_\Delta}
	-  \frac{C^{y}_{VV\Delta}}{C^{y}_\Delta}+
	\frac{C_{3rd}}{|C^{e \gamma}_\Delta| } \right )}
{\left |\frac{C^{e \gamma}_{VV\Delta}}{C^{e \gamma}_\Delta}
	-  \frac{C^{y}_{VV\Delta}}{C^{y}_\Delta}+
	\frac{C_{3rd}}{|C^{e \gamma}_\Delta| }\right |} \,,
\end{eqnarray}
which is estimated to be ${\cal O}(10^{-4})$, as can be confirmed in Fig.\,\ref{fig:eEDM123}.
The predicted  electron EDM is below the experimental upper-bound
after taking account of  the upper-bound of
 $\mathcal{B}({\mu \to e \gamma})$.
In the next subsection, we  discuss the expectation value of
the electron EDM.

We also present the predicted region of  EDMs of muon and tauon  in Fig.\,\ref{fig:mutauEDM}.
 Those are still far from the present experimental upper-bounds,
 $|{d_\mu} |<1.8 \times 10^{-19} \, \rm  e\,cm $ \cite{Muong-2:2008ebm}
 and $|d_\tau|<1.85\times  10^{-17}{\rm e\,cm}$ \cite{Uno:2022xau}
 as seen in Eqs.\,\eqref{muEDM} and \eqref{tauEDM}.

 \subsection{Expectation of
 	 $\mu\to e\gamma$, $\tau\to \mu\gamma$,  $\tau\to e\gamma$ 
 	 and electron EDM}
  \label{frequency}

\begin{figure}[t]
	\begin{minipage}[]{0.47\linewidth}
		\includegraphics[{width=\linewidth}]{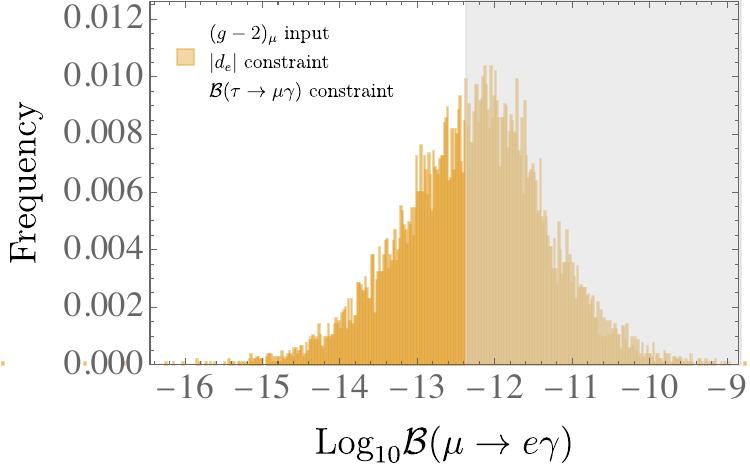}
		\caption{The frequency distribution of the predicted
				$\mathcal{B}({\mu \to e \gamma})$ 
		with  imposing the upper-bounds of $\tau\to \mu\gamma$ decay and
		the electron EDM. The grey region is
			 excluded by the experimental data of $\mathcal{B}({\mu \to e \gamma})$.}
		\label{fig:H-mue}
	\end{minipage}
	\hspace{5mm}
	\begin{minipage}[]{0.47\linewidth}
		\includegraphics[{width=\linewidth}]{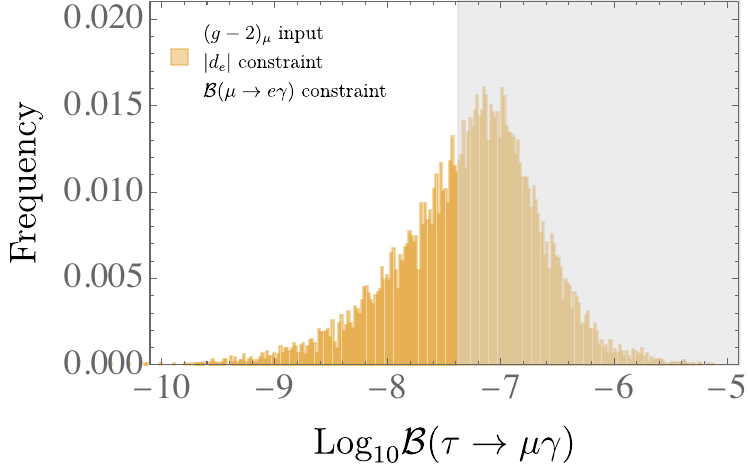}
	\caption{The frequency distribution of the predicted
		$\mathcal{B}({\tau \to \mu \gamma})$ 
		with  imposing the upper-bound of the $\mu\to e\gamma$ decay and
		the electron EDM. The grey region is
		excluded by the experimental data of $\mathcal{B}({\tau \to \mu \gamma})$.}
		\label{fig:H-taumu}
	\end{minipage}
	\label{mue}
\end{figure}

Since we have performed a random scan in the linear space
of $s_e$ and $\epsilon_{\ell}$ in $[0.01-0.1]$ as given in  Eq.\,\eqref{s-ep},
we can show the frequency distribution of the predicted
LFV decays and the electron EDM.
Indeed, the frequency distribution of the predicted
$\mathcal{B}({\mu \to e \gamma})$ by 
imposing  the constraints of the upper-bound on  $\mathcal{B}(\tau\to \mu\gamma)$  and
the electron EDM is shown in Fig.\,\ref{fig:H-mue}. The frequency is normalized so that the total sum is $1$.
 The grey region has already been excluded by the experimental upper-bound on 
  $\mathcal{B}({\mu \to e \gamma})$, 
  and the maximal peak is almost entirely within the gray region.
 Since the future  sensitivity at  the  MEG II experiment is 
   $\mathcal{B}({\mu \to e \gamma})=6\times 10^{-14}$ \cite{MEGII:2018kmf},
 we  expect the observation of  $\mu \to e \gamma$ decay in the near future.
 
  In  Fig.\,\ref{fig:H-taumu}, we plot the frequency distribution of the predicted
 $\mathcal{B}({\tau \to \mu \gamma})$ with imposing  constraints of the upper-bound of  $\mathcal{B}(\mu\to e\gamma)$  and the electron EDM.
  The peak is also within the  grey region.
 We  also expect to observe $\tau \to \mu \gamma$ decay in the near future.

\begin{figure}[H]
	\begin{minipage}[]{0.47\linewidth}
		\includegraphics[{width=\linewidth}]{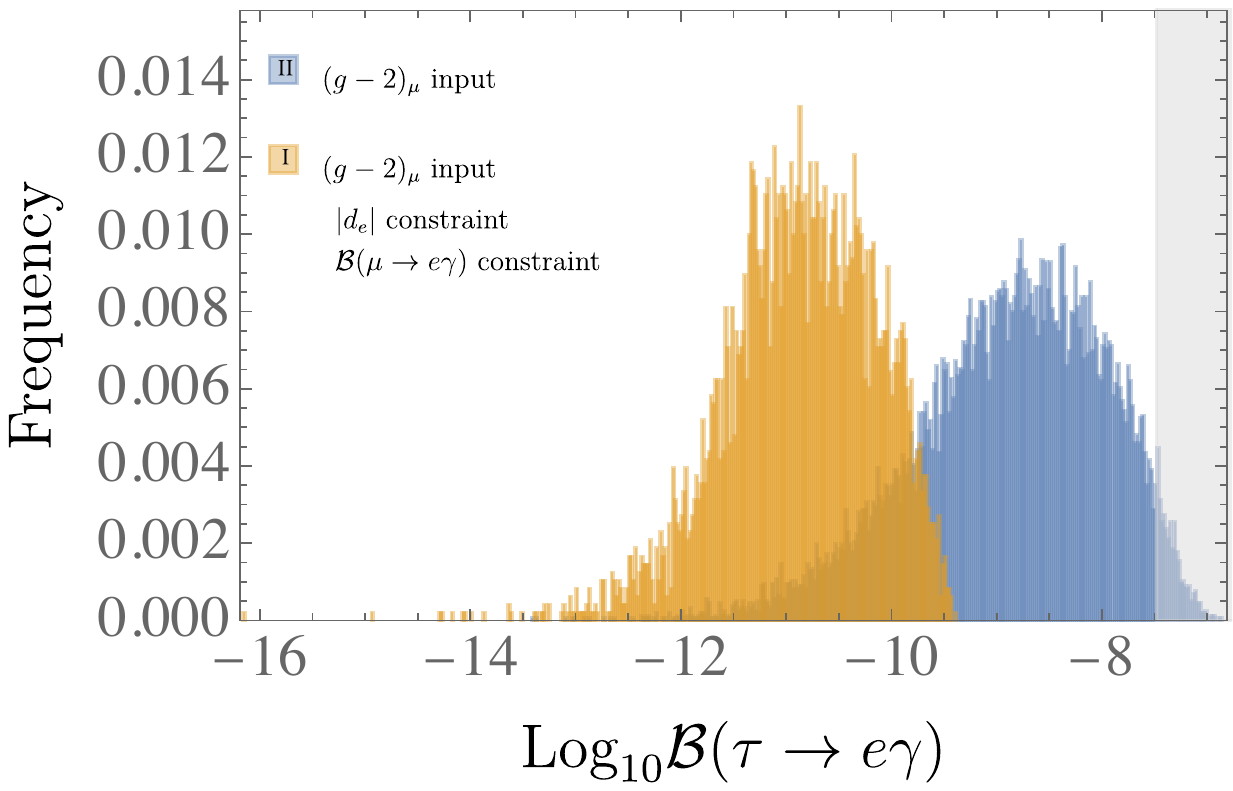}
		\caption{The orange I (blue II) frequency distribution of the predicted
			$\mathcal{B}({\tau \to e\gamma})$ 
			with (without)  imposing the upper-bounds of
			 the $\mu\to e\gamma$ decay,
			$\tau\to \mu\gamma$ decay and the electron EDM.
			 The grey region is
			excluded by the experimental data of	
			$\mathcal{B}({\tau \to e\gamma})$. }
		\label{fig:H-taue}
	\end{minipage}
	\hspace{5mm}
	\begin{minipage}[]{0.47\linewidth}
				\vspace{-6mm}
			\includegraphics[{width=\linewidth}]{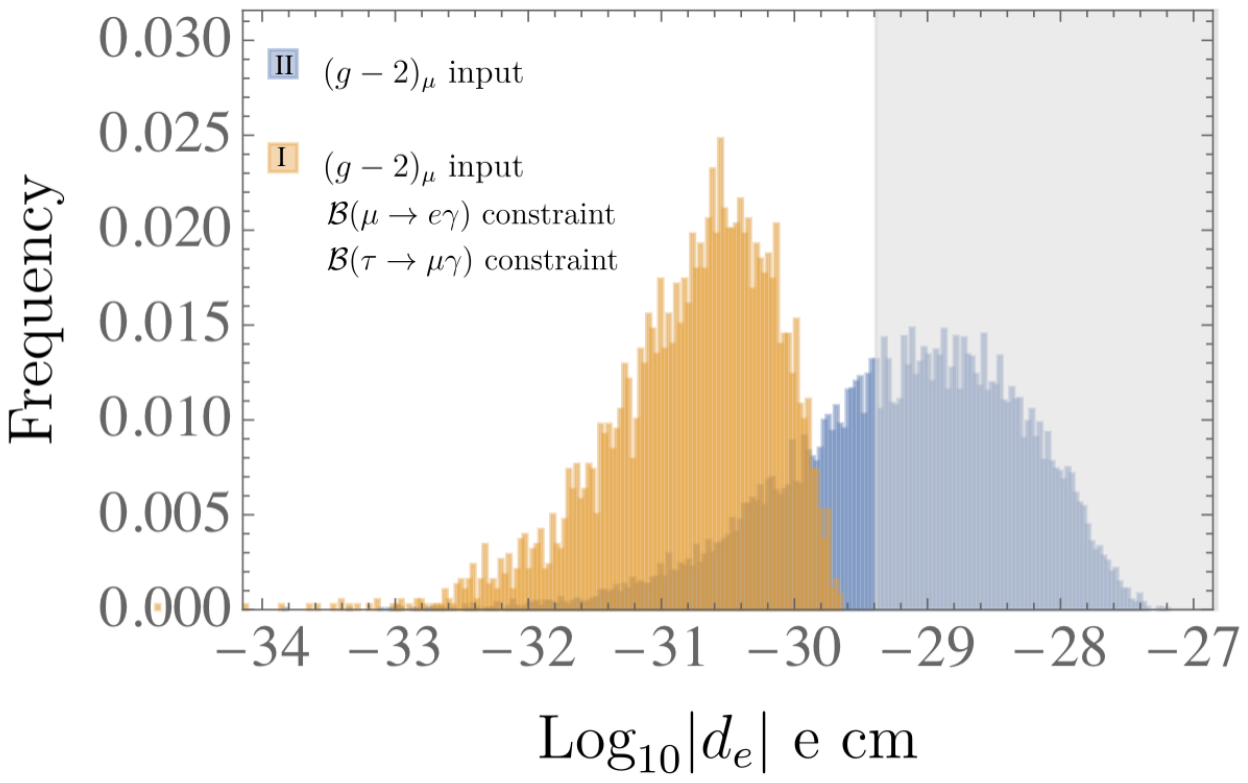}
	\caption{The orange I  (blue II) frequency distribution of the predicted
	electron EDM\,$[\rm e\,cm]$ with (without)  imposing the $\mu\to e\gamma$ decay and
		$\tau\to \mu\gamma$ decay. The grey region is
		excluded by the experimental data of the electron EDM.}
		\label{fig:H-eEDM}
	\end{minipage}
	\label{mue}
\end{figure}

  In  Fig.\,\ref{fig:H-taue}, we plot the frequency distribution of the predicted
  $\mathcal{B}({\tau \to e \gamma})$ with (without)
  imposing  the upper-bounds of  $\mu\to e\gamma$, $\tau\to \mu\gamma$  and the electron EDM with the color of orange I (blue  II).
  Due to these constraints,
  the predicted $\mathcal{B}({\tau \to e\gamma})$ is far from the current experimental upper-bound and is smaller than three orders of magnitude.

Finally, the frequency distribution of the predicted electron EDM with (without)
imposing the constrains of  $\mu\to e\gamma$ and  $\tau\to \mu\gamma$   
is plotted in orange I (blue II) in Fig.\,\ref{fig:H-eEDM}.
Due to these constraints, the peak appears around $|d_e|\simeq 3\times 10^{-31}$e\,cm.
Since the future sensitivity
at ACME $\rm I\hskip -0.04cm I\hskip -0.04cm I$ is expected to be
$|d_e|<3\times10^{-31}\,\text{e\,cm}$ \cite{Kara:2012ay,ACMEIII},
the electron EDM will be possibly  observed in the near future.

\section{Summary and discussions}
\label{summary}

We have studied  the interplay of  NP among
the  lepton magnetic moments,  LFV  and the electron  EDM  in light of recent data of the muon $(g-2)_\mu$.
The NP is discussed in the leptonic dipole operator
with the  {$U(2)_{L_L} \otimes U(2)_{E_R}$} flavor symmetry of the charged leptons,
where  possible CP violating phases of the  three family space are taken into account.
It is remarked that the third-family 
contributes significantly  to  the $\mu \to e\gamma$ decay and the electron EDM.
Indeed, the $\mu \to e\gamma$ process is not suppressed
even if the contribution of  the  first- and second-family vanishes. 
The electron EDM is also predicted to be rather large
due to the CP violating phase in the third-family
even if the  CP phase of the  first- and second-family  is zero.

In addition,
the  electron and tauon  $(g-2)_{e,\tau}$ and the EDM of the muon are discussed,
as well as LFV processes $\tau \to  \mu \gamma$ and  $\tau \to e \gamma$.
In particular, 
the  $\tau_R \to \mu_L \gamma$ process is predicted 
to be close to the experimental upper-bound due to
the  $\mu_L$-$\tau_L$ mixing of ${\cal O}(\epsilon_\ell)$.
This decay will be possibly observed in Belle II experiment in the near future  \cite{Belle-II:2018jsg,DeYtaHernandez:2022bhd}.
The $\tau_R \to e_L \gamma$ decay is not so suppressed
due to the  $e_L$-$\mu_L$ mixing of ${\cal O}(s_e)$ together with
the $\mu_L$-$\tau_L$ mixing.
On the other hand,  the $\mu_L \to e_R \gamma$, $\tau_L \to \mu_R \gamma$
  and $\tau_L \to e_R \gamma$ decays are suppressed.
The angular distribution with respect to the muon polarization can be distinguished between $\mu_R \to e_L\gamma$ and $\mu_L \to e_R\gamma$.
The frequency distributions of the expected values of LFV decays and the electron EDM are presented.
The electron EDM  will be possibly observed  as well as
the $\mu \to e \gamma$ and $\tau \to \mu \gamma$ decays
 in the near future.

Commenting on the numerical results, the predictions for the LFV processes and EDM depend on the scan range of $s_\ell$ and $\epsilon_\ell$ in Eq.\eqref{s-ep}, which are the most natural choices for similar treatments of the quark and lepton sectors. The smaller the values of these two parameters, the smaller the predictions for the LFV process and the EDM. On the other hand, the NP effects of $(g-2)_\mu$ is almost independent of these two parameters.
   
In the numerical analyses, 
the NP effect is taken as $\Delta a_\mu =  249 \times 10^{-11}$.
If $\Delta a_\mu$ is significantly lower, less than or of order $1\,\sigma$,
we obtain somewhat  different predictions for the interplay among  the muon $(g-2)_\mu$, the electron EDM and  
the  $\mu\to e\gamma$ decay.
For example, if we take the input $\Delta a_\mu \simeq 50 \times 10^{-11}$, which is of order $1\,\sigma$, 
the predicted  LFV branching ratios are reduced  by  $1/25$
while  predictions of  $(g-2)_{e,\tau}$ and the predicted EDM are reduced  by $1/5$.
To make the numerical results of the NP contribution to LFV and EDM reliable, 
the precise value of the SM prediction of HVP is needed.


\vspace{0.5 cm}
\noindent
{\large\bf Acknowledgement}\\
This work was supported by  JSPS KAKENHI Grant Number JP21K13923 (KY).

\appendix
\section*{Appendix}

\section{Experimental constraints on the dipole operators}
\label{appen:exp}
From the experimental data of  
the muon $(g-2)_\mu$ and ${\mu \to e\gamma}$,
Ref.\,\cite{Isidori:2021gqe} gave the  constraints on the dipole operators.
We summarize briefly  them on the dipole operators in Eq.\,\eqref{dipole-operators}.
Below the scale of electroweak symmetry breaking,
the leptonic dipole operators are given as:
\begin{align}
\mathcal{O}_{\underset{rs}{e\gamma}}
= \frac{v }{\sqrt{2}}  \overline{e}_{L_r}  \sigma^{\mu\nu} e_{R_s} F_{\mu\nu}\,,
\label{eq:dipoledef}
\end{align}
where $\{r,s\}$ are  {flavor} indices $e,\mu,\tau$ and $F_{\mu\nu}$ is the electromagnetic field strength tensor.
The effective Lagrangian is
\begin{align}
\mathcal{L}_{\rm dipole}=
\frac{1}{\Lambda^2}\,\left (
\CL{e\gamma}{rs}^\prime\mathcal{O}_{\underset{rs}{e\gamma}}
+\CL{e\gamma}{rs}^\prime\mathcal{O}_{\underset{rs}{e\gamma}}
\right )
\,,
\end{align}
where $\Lambda$ is a certain mass scale of NP  in the effective theory. 
The corresponding Wilson coefficient $\CL{e\gamma}{rs}^\prime$
is denoted in the mass-eigenstate basis of leptons.

The tree-level expression for  $\Delta a_\mu$ in terms of the Wilson coefficient of the dipole operator is 
\begin{align}
\Delta a_{\mu} &= \frac{4 m_{\mu}}{e}   \frac{v}{\sqrt 2} \,\frac{1}{\Lambda^2}\text{Re} \, [\CL{e\gamma}{\mu\mu}^\prime] \,,
\label{eq:magnetic-moment}
\end{align}
where
$v\approx 246$~GeV.
Let us input the value
\begin{align}
\Delta a_\mu = 249 \times 10^{-11}~\,,
\label{appen-muon-input}
\end{align}
then, we obtain the Willson coefficient as:
\begin{align}
\frac{1}{\Lambda^2}\text{Re}\  [\CL{e\gamma}{\mu\mu}^\prime]
= 1.0 \times 10^{-5} \, \mathrm{TeV}^{-2} \,, 
\label{Cmumuappen}
\end{align}
where  $e\simeq 0.3028$ is put in the natural unit.

The tree-level expression of a  radiative LFV rate in terms of the Wilson coefficients  is 
\begin{align}
\mathcal{B}(\ell_r \to \ell_s \gamma) = \frac{m_{\ell_r}^3 v^2}{8 \pi \Gamma_{\ell_r}} \frac{1}{\Lambda^4}\left(|\CL{e\gamma}{rs}'|^2 + |\CL{e\gamma}{sr}'|^2\right) \, .
\label{eq:Branching-ratio_lepton-decay}
\end{align}
Taking the experimental bound $\mathcal{B}\!\brackets{\mu^+ \to e^+ \gamma} < 4.2 \times 10^{-13}$~(90\%~C.L.) 
obtained by the MEG experiment~\cite{TheMEG:2016wtm} 
in Eq.\,\eqref{LFV-input},
we obtain  the upper-bound of the Wilson coefficient as:
\begin{align}
\frac{1}{\Lambda^2}|\CL{e\gamma}{e\mu(\mu e)}^\prime| <  2.1 \times 10^{-10} \, \mathrm{TeV}^{-2} \, .
\label{eq:bound_C_egamma_12-appen}
\end{align}

On the other hand,
by taking the following  experimental upper-bound of the branching ratios, $\mathcal{B}\!\brackets{\tau\to\mu+ \gamma} < 4.2 \times 10^{-8}$ and 
$\mathcal{B}\!\brackets{\tau\to e \gamma} < 3.3 \times 10^{-8}$
\cite{BaBar:2009hkt,Belle:2021ysv},
we obtain  the upper-bound of the Wilson coefficient as:
\begin{align}
\frac{1}{\Lambda^2}|\CL{e\gamma}{\mu\tau(\tau\mu)}^\prime| <  2.65 \times 10^{-6} \, \mathrm{TeV}^{-2} \,, \qquad 
\frac{1}{\Lambda^2}|\CL{e\gamma}{e\tau(\tau e)}^\prime| <  2.35 \times 10^{-6} \, \mathrm{TeV}^{-2}  \,,
\label{eq:bound_C_egamma_23-13-appen}
\end{align}
respectively.

\section{Explicit forms of $3\times 3$ matrices for $YY^\dagger$ and $Y^\dagger Y$}
\label{YLLRR}
Here, we present the  $3\times 3$ matrices for 
$YY^\dagger$ and $Y^\dagger Y$ by using Eq.\,\eqref{U2y} explicitly.
We can obtain the left (right)-handed mixing angles and phases by using
these Hermitian matrices. 
The left-handed mixing is obtained by diagonalizing  $YY^\dagger$,
which is: 
{\tiny{
	\begin{align}
	&YY^\dagger=  Y_0^2 \, \times \nonumber\\
	&\begin{pmatrix}
	{C^y_{\Delta}}^2 (c_e^2 {\delta'_e}^2+s_e^2 \delta_e^2) &
	C^y_{\Delta}(C^{y\,*}_{\Delta}+C^{y\,*}_{VV\Delta}\epsilon^2_{\ell}) 
	({\delta'_e}^2-{\delta_e}^2) c_es_e& 
		C^{y}_{\Delta}C^{y\,*}_{V\Delta}({\delta'_e}^2-{\delta_e}^2) c_es_e\epsilon_{\ell}\\
	C^{y\,*}_{\Delta}(C^{y}_{\Delta}+C^{y}_{VV\Delta}\epsilon^2_{\ell}) 
({\delta'_e}^2-{\delta_e}^2) c_es_e &
	|(C^{y}_{\Delta}+C^{y}_{VV\Delta}\epsilon^2_{\ell})|^2
	(s_e^2 {\delta'_e}^2+c_e^2 \delta_e^2)+|C^{y}_{V}|^2\epsilon^2_{\ell}
	&	[ C^{y\,*}_{V\Delta}
	(C^{y}_{\Delta}+C^{y}_{VV\Delta}\epsilon^2_{\ell}) (s_e^2 {\delta'_e}^2+c_e^2 \delta_e^2)+C^{y}_V C^{{y\,*}}]\, \epsilon_{\ell}
	\\
	C^{y\,*}_{\Delta}C^{y}_{V\Delta}({\delta'_e}^2-{\delta_e}^2) c_es_e\epsilon_{\ell}&
	[ C^{y}_{V\Delta}
	(C^{y\,*}_{\Delta}+C^{y\,*}_{VV\Delta}\epsilon^2_{\ell}) (s_e^2 {\delta'_e}^2+c_e^2 \delta_e^2)+C^{y\,*}_V C^{y}]\, \epsilon_{\ell}&
	|C^y|^2+(s_e^2 {\delta'_e}^2+c_e^2 \delta_e^2)|C^{y}_{V\Delta}|^2
	\end{pmatrix}. 
	\label{Y2LL}
	\end{align}
}}
On the other hand, the right-handed mixing is obtained
by diagonalizing  $Y^\dagger Y$,
which is: 
{\tiny{
	\begin{align}
	&Y^\dagger Y= Y_0^2 \, \times \nonumber\\
	&\begin{pmatrix}
	[{C^y_{\Delta}}^2c_e^2+
	(|C^{y}_{\Delta}+C^{y}_{VV\Delta}\epsilon^2_{\ell}|^2+|C^{y}_{V\Delta}|^2\epsilon^2_{\ell})s_e^2] {\delta'_e}^2  &
	(2{\rm Re}{C^y_{\Delta}}+
	|C^{y}_{VV\Delta}|^2\epsilon^2_{\ell} +|C^{y}_{V\Delta}|^2)
	\delta'_e \delta_e  \epsilon^2_{\ell} c_e s_e & 
		(C^{y\,*}_{\Delta}C^{y}_V+ C^{y\,*}_{V}C^{y}+C^{y\,*}_{VV\Delta}C^{y}_V\epsilon^2_{\ell})
	\delta'_e  \epsilon_{\ell} s_e\\
	(2{\rm Re}{C^{y}_{\Delta}}+
	|C^{y}_{VV\Delta}|^2\epsilon^2_{\ell} +|C^{y}_{V\Delta}|^2)
	\delta'_e \delta_e  \epsilon^2_{\ell} c_e s_e&
	[{C^y_{\Delta}}^2 s_e^2+
	(|C^{y}_{\Delta}+C^{y}_{VV\Delta}\epsilon^2_{\ell}|^2+|C^{y}_{V\Delta}|^2\epsilon^2_{\ell})c_e^2] {\delta_e}^2 &
	(C^{y\,*}_{\Delta}C^{y}_V+ C^{y\,*}_{V\Delta}C^{y}+C^{y\,*}_{VV\Delta}C^{y}_V\epsilon^2_{\ell})
\delta_e  \epsilon_{\ell} c_e
	\\
	(C^{y}_{\Delta}C^{y\,*}_V+ C^{y}_{V}C^{y\,*}+C^{y}_{VV\Delta}C^{y\,*}_V\epsilon^2_{\ell})
	\delta'_e  \epsilon_{\ell} s_e&
	(C^{y}_{\Delta}C^{y\,*}_V+ C^{y}_{V\Delta}C^{y\,*}+C^{y}_{VV\Delta}C^{y\,*}_V\epsilon^2_{\ell})
	\delta_e  \epsilon_{\ell} c_e&
	|C^y|^2+|C^{y}_{V}|^2 \epsilon^2_{\ell}
	\end{pmatrix}.
	\label{Y2RR}
	\end{align}
}}



\begin{thebibliography}{99}
	
	\bibitem{Muong-2:2023cdq}
	D.~P.~Aguillard \textit{et al.} [Muon g-2],
	[arXiv:2308.06230 [hep-ex]].
	
	\bibitem{Muong-2:2021ojo}
	B.~Abi \textit{et al.} [Muon g-2],
	Phys. Rev. Lett. \textbf{126} (2021) no.14, 141801
	[arXiv:2104.03281 [hep-ex]].
	
	\bibitem{Muong-2:2006rrc}
	G.~W.~Bennett \textit{et al.} [Muon g-2],
	Phys. Rev. D \textbf{73} (2006), 072003
	[arXiv:hep-ex/0602035 [hep-ex]].
	
	\bibitem{Aoyama:2020ynm}
	T.~Aoyama, N.~Asmussen, M.~Benayoun, J.~Bijnens, T.~Blum, M.~Bruno, I.~Caprini, C.~M.~Carloni Calame, M.~C\`e and G.~Colangelo, \textit{et al.}
	Phys. Rept. \textbf{887} (2020), 1-166
	[arXiv:2006.04822 [hep-ph]].
	
	\bibitem{Jegerlehner:2017gek}
	F.~Jegerlehner,
	Springer Tracts Mod. Phys. \textbf{274} (2017), pp.1-693.
	
	\bibitem{Colangelo:2018mtw}
	G.~Colangelo, M.~Hoferichter and P.~Stoffer,
	JHEP \textbf{02} (2019), 006
	[arXiv:1810.00007 [hep-ph]].
	
	\bibitem{Hoferichter:2019mqg}
	M.~Hoferichter, B.~L.~Hoid and B.~Kubis,
	JHEP \textbf{08} (2019), 137
	[arXiv:1907.01556 [hep-ph]].
	
	\bibitem{Davier:2019can}
	M.~Davier, A.~Hoecker, B.~Malaescu and Z.~Zhang,
	Eur. Phys. J. C \textbf{80} (2020) no.3, 241
	[erratum: Eur. Phys. J. C \textbf{80} (2020) no.5, 410]
	[arXiv:1908.00921 [hep-ph]].
	
	\bibitem{Keshavarzi:2019abf}
	A.~Keshavarzi, D.~Nomura and T.~Teubner,
	Phys. Rev. D \textbf{101} (2020) no.1, 014029
	[arXiv:1911.00367 [hep-ph]].
	
	\bibitem{Hoid:2020xjs}
	B.~L.~Hoid, M.~Hoferichter and B.~Kubis,
	Eur. Phys. J. C \textbf{80} (2020) no.10, 988
	[arXiv:2007.12696 [hep-ph]].
	
	\bibitem{Czarnecki:2002nt}
	A.~Czarnecki, W.~J.~Marciano and A.~Vainshtein,
	Phys. Rev. D \textbf{67} (2003), 073006
	[erratum: Phys. Rev. D \textbf{73} (2006), 119901]
	[arXiv:hep-ph/0212229 [hep-ph]].
	
	\bibitem{Melnikov:2003xd}
	K.~Melnikov and A.~Vainshtein,
	Phys. Rev. D \textbf{70} (2004), 113006
	[arXiv:hep-ph/0312226 [hep-ph]].
	
	\bibitem{Aoyama:2012wk}
	T.~Aoyama, M.~Hayakawa, T.~Kinoshita and M.~Nio,
	Phys. Rev. Lett. \textbf{109} (2012), 111808
	[arXiv:1205.5370 [hep-ph]].
	
	\bibitem{Gnendiger:2013pva}
	C.~Gnendiger, D.~St\"ockinger and H.~St\"ockinger-Kim,
	Phys. Rev. D \textbf{88} (2013), 053005
	[arXiv:1306.5546 [hep-ph]].
	
	
	
	\bibitem{CMD-3:2023alj}
	F.~V.~Ignatov \textit{et al.} [CMD-3],
	[arXiv:2302.08834 [hep-ex]].

	
	\bibitem{Borsanyi:2020mff}
	S.~Borsanyi, Z.~Fodor, J.~N.~Guenther, C.~Hoelbling, S.~D.~Katz, L.~Lellouch, T.~Lippert, K.~Miura, L.~Parato and K.~K.~Szabo, \textit{et al.}
	Nature \textbf{593} (2021) no.7857, 51-55
	[arXiv:2002.12347 [hep-lat]].

\bibitem{Roussy:2022cmp}
T.~S.~Roussy, L.~Caldwell, T.~Wright, W.~B.~Cairncross, Y.~Shagam, K.~B.~Ng, N.~Schlossberger, S.~Y.~Park, A.~Wang and J.~Ye, \textit{et al.}
Science \textbf{381} (2023) no.6653, adg4084
[arXiv:2212.11841 [physics.atom-ph]].

	
	\bibitem{Andreev:2018ayy}
	V.~Andreev \textit{et al.} [ACME],
	Nature \textbf{562} (2018) no.7727, 355-360.
	
	\bibitem{Kara:2012ay}
	D.~M.~Kara, I.~J.~Smallman, J.~J.~Hudson, B.~E.~Sauer, M.~R.~Tarbutt and E.~A.~Hinds,
	New J. Phys. \textbf{14} (2012), 103051
	[arXiv:1208.4507 [physics.atom-ph]].
	
	
	\bibitem{ACMEIII}
	J. Doyle,
	``Search for the Electric Dipole Moment of the Electron with Thorium Monoxide - The ACME Experiment.''Talk at the KITP, September 2016.
	

	
	
	\bibitem{Muong-2:2008ebm}
	G.~W.~Bennett \textit{et al.} [Muon (g-2)],
	Phys. Rev. D \textbf{80} (2009), 052008
	[arXiv:0811.1207 [hep-ex]].
	

\bibitem{Belle:2002nla}
K.~Inami \textit{et al.} [Belle],
Phys. Lett. B \textbf{551} (2003), 16-26
[arXiv:hep-ex/0210066 [hep-ex]].


\bibitem{Bernreuther:2021elu}
W.~Bernreuther, L.~Chen and O.~Nachtmann,
Phys. Rev. D \textbf{103} (2021) no.9, 096011
[arXiv:2101.08071 [hep-ph]].

\bibitem{Uno:2022xau}
K.~Uno,
PoS \textbf{ICHEP2022} (2022), 721.

\bibitem{TheMEG:2016wtm}
A.~M.~Baldini \textit{et al.} [MEG],
Eur. Phys. J. C \textbf{76} (2016) no.8, 434
[arXiv:1605.05081 [hep-ex]].
\bibitem{BaBar:2009hkt}
B.~Aubert \textit{et al.} [BaBar],
Phys. Rev. Lett. \textbf{104} (2010), 021802
[arXiv:0908.2381 [hep-ex]].
\bibitem{Belle:2021ysv}
A.~Abdesselam \textit{et al.} [Belle],
JHEP \textbf{10} (2021), 19
[arXiv:2103.12994 [hep-ex]].	

	
	
	\bibitem{Buchmuller:1985jz}
	W.~Buchmuller and D.~Wyler,
	Nucl. Phys. B \textbf{268} (1986), 621-653.
	
	\bibitem{Grzadkowski:2010es}
	B.~Grzadkowski, M.~Iskrzynski, M.~Misiak and J.~Rosiek,
	JHEP \textbf{10} (2010), 085
	[arXiv:1008.4884 [hep-ph]].
	
	
	\bibitem{Alonso:2013hga}
	R.~Alonso, E.~E.~Jenkins, A.~V.~Manohar and M.~Trott,
	JHEP \textbf{04} (2014), 159
	[arXiv:1312.2014 [hep-ph]].

	\bibitem{Panico:2018hal}
	G.~Panico, A.~Pomarol and M.~Riembau,
	JHEP \textbf{04} (2019), 090
	[arXiv:1810.09413 [hep-ph]].
	
	\bibitem{Aebischer:2021uvt}
	J.~Aebischer, W.~Dekens, E.~E.~Jenkins, A.~V.~Manohar, D.~Sengupta and P.~Stoffer,
	JHEP \textbf{07} (2021), 107
	[arXiv:2102.08954 [hep-ph]].
	
	
	\bibitem{Allwicher:2021rtd}
	L.~Allwicher, P.~Arnan, D.~Barducci and M.~Nardecchia,
	JHEP \textbf{10} (2021), 129
	[arXiv:2108.00013 [hep-ph]].
	
	
  \bibitem{Kley:2021yhn}
  J.~Kley, T.~Theil, E.~Venturini and A.~Weiler,
  Eur. Phys. J. C \textbf{82} (2022) no.10, 926
  [arXiv:2109.15085 [hep-ph]].
	
	
	\bibitem{Isidori:2021gqe}
	G.~Isidori, J.~Pag\`es and F.~Wilsch,
	JHEP \textbf{03} (2022), 011
	[arXiv:2111.13724 [hep-ph]].
	
	
	

	
	\bibitem{Barbieri:2011ci}
	R.~Barbieri, G.~Isidori, J.~Jones-Perez, P.~Lodone and D.~M.~Straub,
	Eur. Phys. J. C \textbf{71} (2011), 1725
	[arXiv:1105.2296 [hep-ph]].
	
	\bibitem{Barbieri:2012uh}
	R.~Barbieri, D.~Buttazzo, F.~Sala and D.~M.~Straub,
	JHEP \textbf{07} (2012), 181
	[arXiv:1203.4218 [hep-ph]].
	

	
	\bibitem{Blankenburg:2012nx}
	G.~Blankenburg, G.~Isidori and J.~Jones-Perez,
	Eur. Phys. J. C \textbf{72} (2012), 2126
	[arXiv:1204.0688 [hep-ph]].
	
	\bibitem{Chivukula:1987py}
	R.~S.~Chivukula and H.~Georgi,
	Phys. Lett. B \textbf{188} (1987), 99-104.
	
	\bibitem{DAmbrosio:2002vsn}
	G.~D'Ambrosio, G.~F.~Giudice, G.~Isidori and A.~Strumia,
	Nucl. Phys. B \textbf{645} (2002), 155-187
	[arXiv:hep-ph/0207036 [hep-ph]].
	
	\bibitem{Fuentes-Martin:2019mun}
	J.~Fuentes-Mart\'\i{}n, G.~Isidori, J.~Pag\`es and K.~Yamamoto,
	Phys. Lett. B \textbf{800} (2020), 135080
	[arXiv:1909.02519 [hep-ph]].
	
	\bibitem{Faroughy:2020ina}
	D.~A.~Faroughy, G.~Isidori, F.~Wilsch and K.~Yamamoto,
	JHEP \textbf{08} (2020), 166
	[arXiv:2005.05366 [hep-ph]].
	
	\bibitem{Kobayashi:2021pav}
	T.~Kobayashi, H.~Otsuka, M.~Tanimoto and K.~Yamamoto,
	Phys. Rev. D \textbf{105} (2022) no.5, 055022
	[arXiv:2112.00493 [hep-ph]].
\bibitem{Kobayashi:2022jvy}
T.~Kobayashi, H.~Otsuka, M.~Tanimoto and K.~Yamamoto,
JHEP \textbf{08} (2022), 013
[arXiv:2204.12325 [hep-ph]].

\bibitem{Kobayashi:2021uam}
T.~Kobayashi and H.~Otsuka,
Eur. Phys. J. C \textbf{82} (2022) no.1, 25
[arXiv:2108.02700 [hep-ph]].

\bibitem{Calibbi:2021qto}
L.~Calibbi, M.~L.~L\'opez-Ib\'a\~nez, A.~Melis and O.~Vives,
Eur. Phys. J. C \textbf{81} (2021) no.10, 929
[arXiv:2104.03296 [hep-ph]].



\bibitem{Altarelli:2010gt}
G.~Altarelli and F.~Feruglio,
Rev. Mod. Phys. \textbf{82} (2010), 2701-2729
[arXiv:1002.0211 [hep-ph]].

\bibitem{Ishimori:2010au}
H.~Ishimori, T.~Kobayashi, H.~Ohki, Y.~Shimizu, H.~Okada and M.~Tanimoto,
Prog. Theor. Phys. Suppl. \textbf{183} (2010), 1-163
[arXiv:1003.3552 [hep-th]].



\bibitem{Ishimori:2012zz}
H.~Ishimori, T.~Kobayashi, H.~Ohki, H.~Okada, Y.~Shimizu and M.~Tanimoto,
Lect. Notes Phys. \textbf{858} (2012), 1-227
doi:10.1007/978-3-642-30805-5

\bibitem{Kobayashi:2022moq}
T.~Kobayashi, H.~Ohki, H.~Okada, Y.~Shimizu and M.~Tanimoto,
Lect.\ Notes Phys.\ {\bf 995} (2022) 1, Springer 
doi:10.1007/978-3-662-64679-3.

\bibitem{Hernandez:2012ra}
D.~Hernandez and A.~Y.~Smirnov,
Phys. Rev. D \textbf{86} (2012), 053014
[arXiv:1204.0445 [hep-ph]].

\bibitem{King:2013eh}
S.~F.~King and C.~Luhn,
Rept. Prog. Phys. \textbf{76} (2013), 056201
[arXiv:1301.1340 [hep-ph]].

\bibitem{King:2014nza} 
S.~F.~King, A.~Merle, S.~Morisi, Y.~Shimizu and M.~Tanimoto,
New J.\ Phys.\  {\bf 16}, 045018 (2014)
[arXiv:1402.4271 [hep-ph]].


\bibitem{Tanimoto:2015nfa}
M.~Tanimoto,
AIP Conf. Proc. \textbf{1666} (2015) no.1, 120002

\bibitem{King:2017guk}
S.~F.~King,
Prog. Part. Nucl. Phys. \textbf{94} (2017), 217-256
[arXiv:1701.04413 [hep-ph]].

\bibitem{Petcov:2017ggy}
S.~T.~Petcov,
Eur.\ Phys.\ J.\ C {\bf 78} (2018) no.9,  709
[arXiv:1711.10806 [hep-ph]].


\bibitem{Feruglio:2019ybq}
F.~Feruglio and A.~Romanino,
Rev. Mod. Phys. \textbf{93} (2021) no.1, 015007
[arXiv:1912.06028 [hep-ph]].



\bibitem{Shafi:2023ksr}
Q.~Shafi, A.~Tiwari and C.~S.~Un,
[arXiv:2308.14682 [hep-ph]].

\bibitem{Buttazzo:2020ibd}
D.~Buttazzo and P.~Paradisi,
Phys. Rev. D \textbf{104} (2021) no.7, 075021
[arXiv:2012.02769 [hep-ph]].


\bibitem{Bordone:2017anc}
M.~Bordone, G.~Isidori and S.~Trifinopoulos,
Phys. Rev. D \textbf{96} (2017) no.1, 015038
[arXiv:1702.07238 [hep-ph]].

\bibitem{Carone:1997qg}
C.~D.~Carone and L.~J.~Hall,
Phys. Rev. D \textbf{56} (1997), 4198-4206
[arXiv:hep-ph/9702430 [hep-ph]].

\bibitem{Tanimoto:1997zw}
M.~Tanimoto,
Phys. Rev. D \textbf{57} (1998), 1983-1986
[arXiv:hep-ph/9706497 [hep-ph]].

\bibitem{Blazek:1999ue}
T.~Blazek, S.~Raby and K.~Tobe,
Phys. Rev. D \textbf{60} (1999), 113001
[arXiv:hep-ph/9903340 [hep-ph]].

\bibitem{Giudice:2012ms}
G.~F.~Giudice, P.~Paradisi and M.~Passera,
JHEP \textbf{11} (2012), 113
[arXiv:1208.6583 [hep-ph]].

%

\bibitem{Hanneke:2008tm}
D.~Hanneke, S.~Fogwell and G.~Gabrielse,
Phys. Rev. Lett. \textbf{100} (2008), 120801
[arXiv:0801.1134 [physics.atom-ph]].

\bibitem{Parker:2018vye}
R.~H.~Parker, C.~Yu, W.~Zhong, B.~Estey and H.~M\"uller,
Science \textbf{360} (2018), 191
[arXiv:1812.04130 [physics.atom-ph]].

\bibitem{Rb:2020}
L. Morel,  Z. Yao, P. Clad\'e, and S. Guellati-Kh\'elifa,
Nature \textbf{588} (2020), no.7836 61-65.	



 
\bibitem{Nishimura:2023wdu}
S.~Nishimura, C.~Miyao and H.~Otsuka,
[arXiv:2304.14176 [hep-ph]].

\bibitem{Okada:1999zk}
Y.~Okada, K.~i.~Okumura and Y.~Shimizu,
Phys. Rev. D \textbf{61} (2000), 094001
[arXiv:hep-ph/9906446 [hep-ph]].

\bibitem{Belle-II:2018jsg}
E.~Kou \textit{et al.} [Belle-II],
PTEP \textbf{2019} (2019) no.12, 123C01
[erratum: PTEP \textbf{2020} (2020) no.2, 029201]
[arXiv:1808.10567 [hep-ex]].

\bibitem{DeYtaHernandez:2022bhd}
A.~De Yta Hern\'andez [Belle II],
PoS \textbf{EPS-HEP2021} (2022), 527.


\bibitem{MEGII:2018kmf}
A.~M.~Baldini \textit{et al.} [MEG II],
Eur. Phys. J. C \textbf{78} (2018) no.5, 380
[arXiv:1801.04688 [physics.ins-det]].






\end{thebibliography}
\end{document}